\documentclass[preprint2]{aastex62}

\submitjournal{ApJS}

\usepackage{amsmath}
\usepackage{amssymb}

\usepackage{savesym}
\savesymbol{tablenum}
\usepackage{siunitx}
\restoresymbol{SIX}{tablenum}

\DeclareSIUnit[]{\hill}{h}
\DeclareSIUnit[]{\deg}{deg}
\DeclareSIUnit[]{\orbits}{orbits}
\DeclareSIUnit[]{\orbit}{orbit}
\DeclareSIUnit[]{\years}{years}
\DeclareSIUnit[]{\freq}{\per\second}
\DeclareSIUnit[]{\viscosity}{\centi\meter\squared\per\second}
\DeclareSIUnit[]{\resolution}{\hill\per pixel}
\newcommand{\bleriot}{Bl\'eriot}

\begin{document}


\title{Hydrodynamic Simulations of Asymmetric Propeller Structures in Saturn's Rings}

\author{M. Seiler}
\affiliation{Theoretical Physics Group, Institute of Physics and 
  Astronomy, University of Potsdam, Germany}
\author{M. Sei\ss}
\affiliation{Theoretical Physics Group, Institute of Physics and 
  Astronomy, University of Potsdam, Germany}
\author{H. Hoffmann}
\affiliation{Theoretical Physics Group, Institute of Physics and 
  Astronomy, University of Potsdam, Germany}
\author{F. Spahn}
\affiliation{Theoretical Physics Group, Institute of Physics and 
  Astronomy, University of Potsdam, Germany}

\correspondingauthor{M. Seiler}
\email{miseiler@uni-potsdam.de}

\begin{abstract}

The observation of the non-Keplerian behavior of propeller structures 
in Saturn's outer A ring \citep{Tiscareno2010ApJL,Seiler2017ApJL,Spahn2018Book} 
raises the question, how the propeller responds to the wandering of the central 
embedded moonlet. Here, we study numerically how the induced propeller is 
changing for a librating moonlet. It turns out, that the libration of 
the moonlet induces an asymmetry in the structural imprint of the propeller, 
where the asymmetry is depending on the moonlet's libration period and 
amplitude. Further, we study the dependence of the asymmetry on the libration 
period and amplitude for a moonlet with \SI{400}{\meter} Hill radius, which is 
located in the outer A ring. In this way, we are able to apply our findings to 
the largest found propeller structures -- such as \bleriot{} -- which are 
expected to be of similar size. For \bleriot, we can conclude that, supposed 
the moonlet is librating with the largest observed period of \SI{11.1}{\years} 
and an azimuthal amplitude of about \SI{1845}{\kilo\meter} 
\citep{Seiler2017ApJL,Spahn2018Book}, a small assymetry should be measurable 
but depends on the moonlet's libration phase at the observation time. 
The excess motions of the other giant propellers -- such as Santos Dumont and 
Earhart -- have similar amplitudes as \bleriot{} and thus might allow the 
observation of larger asymmetries due to their smaller azimuthal extent. 
This would permit to scan the whole gap structure for asymmetries.

Although the librational model of the moonlet is a simplification, our 
results are a first step towards the development of a consistent model for 
the description of the formation of asymmetric propellers caused by a freely
moving moonlet.

\end{abstract}

\keywords{
hydrodynamics ---
methods: numerical  ---
methods: data analysis ---
planets and satellites: dynamical evolution and stability ---
planets and satellites: rings ---
planets and satellites: individual(Saturn)
}



\section{Introduction} \label{sec:intro}
Planetary rings are one of the most beautiful structures in our 
solar system, which are mainly composed of differently sized icy particles.
The dense main rings of Saturn are composed of particles ranging from 
centimeters to a few meters \citep{Cuzzi2018Book} and beside this ensemble of 
particles, even larger boulders (called \emph{moonlets}) are orbiting within 
the ring environment.
By their strong gravitational interaction with the surrounding ring material, 
these objects try to sweep free a gap around their orbits. Due to the 
different angular speeds of the ring material for varying semi-major axis 
the gap appears inside and outside the moonlet's mean radial position with a 
slight radial shift. Therefore, the inner gap -- due to the higher angular 
speed of the ring material -- overtakes the embedded moonlet while the outer 
one falls behind.
Viscous diffusion of the ring material counteracts this gravitational 
scattering process and thus results in a closing of the opened gap with 
growing azimuthal distance to the embedded moonlet 
\citep{Spahn2000AAP,Sremcevic2002MNRAS,Seiss2005GeoRL,Hoffmann2015Icar}. 
For moonlets larger than a critical radius, its gravity compensates the 
viscous diffusion along the whole circumference of the orbit and thus the 
gap is kept open. For smaller moonlet radii, the gravity of the moonlet does 
not suffice to compensate the viscous diffusion and therefore the gap closes. 
As a result, two partial gaps inside and outside the orbit of the moonlet
decorated with wakes remain and resemble an S-shaped, two-bladed 
\emph{propeller}. Those propellers act as structural insignia to detect the 
embedded moonlet in the dense rings.

After the postulation of their existence in Saturn's ring to explain optical
depth variations in the Voyager data \citep{Henon1981Nat,Lissauer1981bNat}, 
the cameras onboard the spacecraft Cassini finally revealed their presence 
within the dense rings of Saturn \citep{Tiscareno2006Nat,Spahn2006Nat,
Sremcevic2007Nat, Tiscareno2008AJ}.
Meanwhile, different populations of propeller structures inside Saturn's A ring
have been identified in the images. Between \num{127000} and 
\SI{132000}{\kilo\meter} from Saturn's center three \emph{propeller belts} have
been observed, which contain several thousand propellers, all being generated 
by moonlets of radii $\lesssim$ \SI{0.15}{\kilo\meter} \citep{Tiscareno2008AJ}. 
Further outwards, between the Encke and Keeler gap, about 37 larger propellers 
being created by moonlets with radii between $\sim$\SI{0.15}{\kilo\meter} and 
$\lesssim$\SI{1.0}{\kilo\meter} have been found \citep{Tiscareno2010ApJL}. 
Due to their relatively large size, 11 of these \emph{giant-propellers} were 
able to be identified in different subsequent images, allowing the 
reconstruction of their orbital motion. The analysis of their orbital 
evolution revealed a longitudinal deviation from their expected Keplerian 
location -- called \emph{excess motion} 
\citep{Tiscareno2010ApJL,Spahn2018Book}. 

The largest propeller structure is called \bleriot{} and is created by a 
moonlet of about \SI{400}{\meter} in Hill radius \citep{Hoffmann2016DPS,
Seiss2017arXiv}. Its excess motion can be reconstructed by a superposition 
of three sinusoidal harmonics with periods of \num{11.1}, \num{3.7} and 
\SI{2.2}{\years} and amplitudes of \num{1845}, \num{152} and 
\SI{58}{\kilo\meter} with a standard deviation of about 
\SI{17}{\kilo\meter} of the remaining residual \citep{Seiler2017ApJL, 
Spahn2018Book}. 
\bleriot's excess motion resembles the typical librational motion of a 
resonantly perturbed moon \citep{Goldreich1965cMNRAS, Goldreich2003aIcar,
Goldreich2003bIcar,Spitale2006AJ,Cooper2015AJ} and thus supports the 
hypothesis, that this excess motion might result from a resonant or 
near-resonant driving by one or several of the outer satellites.  In 
numeric simulations \footnote[1]{In these simulations a test moonlet, 
which has been placed on the expected orbital position of \bleriot{}, has 
been driven gravitationally by Saturn and 15 of its larger moons.} it has 
been shown, that the collective perturbation by the outer satellites 
Prometheus, Pandora and Mimas is able to induce correct libration 
frequencies to explain the three mode fit, but the induced libration 
amplitudes are by far to small to explain the observations 
\citep{Seiler2017ApJL}. 

Alternatively, a stochastic migration of the embedded moonlet due to 
collisions or density fluctuations in the rings has been proposed as well 
\citep{Crida2010AJ,Rein2010AAP,Pan2010ApJL,Pan2012ApJ,
Pan2012MNRAS,Bromley2013ApJ,Tiscareno2013PSS}. The most promising results have
been obtained by \citet{Rein2010AAP}, who considered a random walk in the
semi-major axis of the moonlet, and in N-Body simulations \citet{Pan2012MNRAS} 
have been able to generate an amplitude in the moonlet's excess motion of
about \SI{300}{\kilo\meter} over a time span of \SI{4}{\years}.
Nevertheless, their resulting amplitude is also too small.

In order to solve this amplitude problem \citet{Seiler2017ApJL} suggested a 
propeller-moonlet interaction model, where the usually assumed
point-symmetry of the propeller structure is broken by a slight
displacement of the central moonlet. This breaking of the point-symmetry 
causes a restoring force to the moonlet, resulting in a harmonic oscillation 
in the moonlet's mean longitude. The libration depends on the physical 
parameters of the propeller and the surrounding ring material. This 
oscillating system has its own eigenfrequency so that for an external 
resonant driving with a frequency, which is sufficiently close to the 
eigenfrequency of the propeller-moonlet oscillator, an originally small 
libration mode can be amplified by several orders of magnitude.

However, the suggested propeller-moonlet interaction model is oversimplifying 
the physical problem in the aspect, that it reduces the gap structures to two 
radially separated rectangular shaped boxes with fixed azimuthal end points, 
where the beginning of the gap follows the moonlet motion. Further, the surface
mass density inside these boxes is assumed to be constant. In reality, 
the gap structure has an azimuthal extent of up to several thousand kilometers 
and the surface mass density within the gap structure is rather a 
decay than constant. Due to the radial separation of the gaps from the moonlet, 
the purturbation caused by the moonlet's motion first needs time to be 
transported along the gap structure, given by the Kepler shear. 
Thus, the perturbation by the moonlet's motion only arrives at the gap ends 
after a certain delay time (hereafter called $T_{gap}$), while the beginning 
of the gap more or less follows the moonlet motion instantaneously. For 
\bleriot{} this delay time would be about $T_{gap}\approx$\SI{0.5}{\years} 
considering a gap length of \SI{6500}{\kilo\meter}, respectively 
\citep{Seiler2017ApJL}. 
As a consequence of the moonlet's wandering and the retardation in the 
reaction of the gap ends, the gap structures are expected to become asymmetric.  
Here, this expectation is tested for a given moonlet libration to answer the
question, whether such an asymmetry might be detectable in the Cassini ISS 
images. Therefore, we search the simulated gap structures for properties which 
allow a direct measurement of the asymmetry. 
In addition, we will study the dependence of the asymmetry on the libration 
amplitude and period.

The plan of this paper is as follows:
First, we give an overview of our hydrodynamic simulation code and how we model 
the moonlet libration (see Section \ref{sec:hydro}). Afterwards, we
present the results of our simulations and show, how the appearance of the
propeller and the different properties of the gap structures change with 
the libration of the moonlet. For this purpose, we directly compare
our results to simulations of a symmetric propeller structure (see Section
\ref{sec:asymmetry}). The propeller-moonlet interaction model predicts, that 
the restoring force by the gap-asymmetry follows the moonlet motion. 
This fact will be checked against our simulation data 
(see Section \ref{sec:ring_gravity}). 
As a next step, we study the dependence of the asymmetry on the libration 
period and amplitude and test, under which conditions the asymmetry becomes 
detectable (see Section \ref{sec:predictions}).
On the basis of our analysis, we will apply our findings to the propeller
\bleriot{} in Section \ref{sec:application_bleriot}. In the end of this paper,
we conclude and discuss our findings (see Section \ref{sec:discussion}).

\section{Hydrodynamic Simulation} \label{sec:hydro}
Our hydrodynamic simulation routine bases on the code, developed by 
\citet{Seiss2017arXiv}. This integration routine will be modified so that 
the moonlet can librate around its mean orbital position within the 
simulation grid. 

We consider a moonlet embedded in the granular environment of Saturn's rings 
with surface mass density $\Sigma$. The evolution of the surface mass density 
is described by the continuity equation

\begin{align}
\partial_t \Sigma + {\bf \nabla} \cdot \Sigma {\bf v}  & = 0 
\label{equ:continuity}
\end{align}

and the momentum balance is quantified by the Navier Stokes equation

\begin{align}
\partial_t \Sigma {\bf v} + \nabla \cdot (\Sigma {\bf v} \circ {\bf v}) 
       & = -\Sigma \, \nabla \cdot  (\Phi_{\rm p} + \Phi_{\rm m}) + {\bf f_{\rm i}}
       - {\bf {\nabla}} \cdot {\sf \hat{P}} \, .
\label{equ:navier_stokes}
\end{align}

Both equations above are presented in the flux conserved form, where the 
tensor product is denoted by the $\circ$ symbol. 
The symbol $\partial_t = \partial/\partial t$ denotes the partial time 
derivative. The gravitational potentials of the central planet and the moonlet 
are described by $\Phi_{\rm p}$ and $\Phi_{\rm m}$.
Further, we choose our simulation's coordinate system to co-rotate with the 
moonlet's mean orbital frequency $\Omega = \sqrt{GM_p/r_0^3}$, where $G$ and $M_p$ 
denote the gravitational constant and the mass of the central planet, and to be 
centered at the moonlet's mean radial position $r_0$. 
Here, we neglect the oblateness of Saturn \footnote[1]{
Including the oblateness of Saturn would result in slightly different orbital 
frequencies and in a precession of the ascending node and the pericenter, which 
will not have a significant impact on the results, since we are only considering 
a small ring area in the close vicinity of the moonlet.}.

In comparison to their radial and longitudinal extent (several hundred thousand
kilometers) the vertical dimension ($\lesssim \SI{100}{\meter}$) of Saturn's 
rings is tiny. Thus, equations \ref{equ:continuity} and \ref{equ:navier_stokes} 
can be reduced to a two-dimensional problem using vertically integrated 
quantities \citep{Spahn2018Book}. Additionally, the moonlets in the outer A 
ring are expected to evolve on nearly circular orbits in the equatorial plane 
of Saturn. 

In comparison to the spatial extent of the rings the propellers are tiny. 
As a consequence, only a small ring area, centered around the moonlet's orbital 
position needs to be considered for our simulations. Therefore, the acting 
forces in the vicinity of the moonlet can be linearized and described by the 
Hill problem \citep{Hill1878AmJM}. Here, we choose $x$ to represent the radial 
distance from the moonlet and $y$ stands for the azimuthal direction. 
As a result, the gravitational potential of the central moonlet is given by

\begin{equation}
  \Phi_m = - \frac{G M_{\rm m}}{\sqrt{x^2+y^2+\epsilon^2}} \; ,
\end{equation}

with $\epsilon$ the smoothing radius which limits the gravitational 
potential in the close vicinity of the moonlet's center. Here, we choose 
$\epsilon=\SI{0.2}{\hill}$. The mass of the moonlet $M_m$ is defined by its 
Hill radius

\begin{equation}
  h =  a_0 \sqrt{ \frac{M_m}{3 M_p} } \, ,
\end{equation}

where $a_0$ denotes the semi-major axis of the moonlet's mean orbit. 

We treat the granular ensemble forming dense rings as a usual (linear) fluid. 
Thus, the pressure tensor ${\sf \hat{P}}$ can be described with the Newtonian 
ansatz as

\begin{equation}
P_{ij} = p \, \delta_{ij} - \Sigma \nu  \left( \frac{\partial v_i}{
  \partial x_j} + \frac{\partial v_j}{\partial x_i} \right)
  + \Sigma \left( \frac{2}{3} \, \nu - \xi \right) \, {\bf 
    {\nabla}} \cdot {\bf v} \,\delta_{ij}
\end{equation}

where $p$, $\nu$ and $\xi$ denote the scalar pressure, the kinematic shear and 
the bulk viscosities, respectively. The dependence of the pressure and the 
viscosities on the local density can be described by the power laws
\citep{Spahn2000Icar}

\begin{eqnarray}
p & = & p_0 \, \left( \frac{\Sigma}{\Sigma_0} \right)^{\alpha} \\
\nu & = & \nu_0 \, \left( \frac{\Sigma}{\Sigma_0} \right)^{\beta} \, .
\end{eqnarray}

Additionally, we set the ratio between shear and bulk viscosity constant

\begin{equation}
\xi = \frac{\xi_0}{\nu_0} \cdot \nu \, .
\end{equation}

In the simplest case, the unperturbed pressure is given by the ideal gas 
relation $p_0=\Sigma_0 \, c_0^2$ at equilibrium, where $c_0$ denotes the 
dispersion velocity. In our simulation scheme, we use an isothermal model and
therefore the dispersion velocity $c_0$ and the related granular temperature 
$T=c_0^2/3$ are constant.

\subsection{Methods}

The simulation area is a rectangular cut out of the ring environment 
of dimensions $(x_{\rm min},x_{\rm max})$ and $(y_{\rm min},y_{\rm max})$,
which is centered around the mean orbital position of the moonlet. 
This box is divided into $N_x \times N_y$ equal-sized cells. The complete set 
of equations is integrated until a steady state is established. 
The advection term is solved with the second-order scheme with 'MinMod' flux 
limiter \citep{Leveque2002Book} in order to better conserve the wake crests. 
The influence of pressure and viscous transport is solved with an explicit 
scheme. At the borders of the calculation regions, the boundary conditions are
chosen such, that the perturbations can flow out of the box freely, while
the inflow is unperturbed. This is especially important at the
azimuthal boundaries, where due to Kepler shear the material is flowing
into the box at $x<0,y=y_{\rm min}$ and $x>0,y=y_{\rm max}$ and flowing
out at $x<0,y=y_{\rm max}$ and $x>0,y=y_{\rm min}$. The influence
of the moonlet is established mainly by its gravity. 

A more detailed overview of the implementation of the integration routine 
and the used numerical methods is given in Section 3 in \citet{Seiss2017arXiv}.

The simulation scheme uses following scaled values: 
$x \rightarrow x/h$, $y\rightarrow y/h$, $t\rightarrow\Omega t \sim t/T$ 
and $\Sigma \rightarrow \Sigma/\Sigma_0$.

Here, we will focus on the asymmetry formation for the giant trans-Encke 
propellers. For this reason, we will consider a \bleriot-sized moonlet with 
\SI{400}{\meter} Hill radius \citep{Hoffmann2016DPS, Seiss2017arXiv}, allowing 
us to predict the asymmetry of its propeller.
Therefore, we will simulate the moonlet at a radial location close to its 
observed orbital position and we will adopt the viscosity accordingly
\citep{Seiss2017arXiv}. The set of used parameters and their values -- if nowhere
else mentioned -- are given in Table \ref{tab:simulate_params}.

\begin{table}
\centering
\begin{tabular}{c|c}
  Parameter     & Value \\ \hline
  $h$           & \SI{400}{\meter} \\
  $\Omega$      & \SI{1.3e-3}{\freq} \\
  $\nu_0$       & \SI{300}{\viscosity} \\
  $c_0$         & \SI{0.5}{\centi\meter\per\second}
\end{tabular}
\caption{\footnotesize Simulation parameters for the hydrodynamic 
simulation. Those parameters are used for the scaling within the integration
routine.}
\label{tab:simulate_params}
\end{table}

The scaled viscosity and sound speed are calculated according to the moonlet size 

\begin{align}
  \tilde{\nu} & = \frac{\nu_0}{h^2 \, \Omega} \; \text{and} \\
  \tilde{c}   & = \frac{c_0}{h \, \Omega} \; . \label{equ:nu_c}
\end{align}

Their value and further parameters used for the simulation are given in Table
\ref{tab:further_params}.

\begin{table}
\centering
\begin{tabular}{c|c} 
  Parameter     & Value \\ \hline
  $\tilde{\nu}$ & \num{1.44 1e-3} \\
  $\tilde{c}$   & \num{4.81 1e-2} \\
  $\alpha$      & \num{1} \\
  $\beta$       & \num{2} \\
  $\tilde{\xi}$         & \num{7} $\tilde{\nu}$ \\
  $x_{\rm min}$       & \SI{-10}{\hill} \\   
  $x_{\rm max}$       & \SI{10}{\hill} \\    
  $y_{\rm min}$       & \SI{-800}{\hill} \\   
  $y_{\rm max}$       & \SI{800}{\hill} \\    
  $N_{x}$       & \num{400} \\   
  $N_{y}$       & \num{6400}  
\end{tabular}
\caption{\footnotesize Further simulation parameters used for the 
hydrodynamic simulation.}
\label{tab:further_params}
\end{table}

\subsection{Modeling the Moonlet Motion}
The observed harmonic excess motion of \bleriot{} requires a periodic 
azimuthal motion of the propeller structure. In the easiest way, this can be 
achieved by a libration of the central moonlet. 
Setting the moonlet on a librational motion around its mean orbital position
further allows to directly test the assumptions of the propeller-moonlet 
interaction model \citep{Seiler2017ApJL}. 

The moonlet will be set on a librational motion with amplitude 
$x_{m,0}$ and period $T_m$ given by

\begin{align}
x_m & = x_{m,0}\cos \left( \frac{2\pi}{T_m} t \right) \; \text{and} \label{equ:moonlet_motion} \\ 
y_m & = -\frac{3}{2} x_{m,0} T_m \sin \left( \frac{2\pi}{T_m} t \right) \, .
\end{align}

Note, that the above presented equations for the moonlet motion are already 
given in the scaled form, where the time $t$ and the libration period $T_m$ 
are given in number of orbits and the directions $x_m$ and $y_m$ are given 
in Hill radii. The moonlet's azimuthal amplitude is motivated by the 
proportionality of $dn/dt \propto da/dt$ and thus it is set by the Kepler shear.

The advantage of the librational moonlet motion lies in the fact, that it 
covers both, libration and migration. The resulting asymmetry for a linearly 
migrating moonlet can be studied considering only part of the moonlet 
libration. In this context, the relative velocity of the perturbed moonlet 
motion to its unperturbed orbit -- which depends on the amplitude 
$x_{m,0}$ and period $T_m$ -- will also set the drift rate of the moonlet.

\section{Visualization of the Asymmetry} \label{sec:asymmetry}
For the presentation of our simulation results, we will define and analyze 
different characteristics of the propeller shape and directly compare the
results for a symmetric and asymmetric propeller. In this way, we show, 
how the asymmetry of the propeller can be characterized.
Our results have been obtained for a moonlet with \SI{400}{\meter} Hill radius. 
For the asymmetric propeller -- if nowhere else mentioned --the moonlet has 
been forced to around its mean orbital position with a radial amplitude of 
$x_{m,0} = \SI{0.5}{\hill}$ and period of $T_{m} = \num{80}$ orbital periods.

\subsection{Density Profile}
\begin{figure*}
\centering
\includegraphics[width=0.49\textwidth]{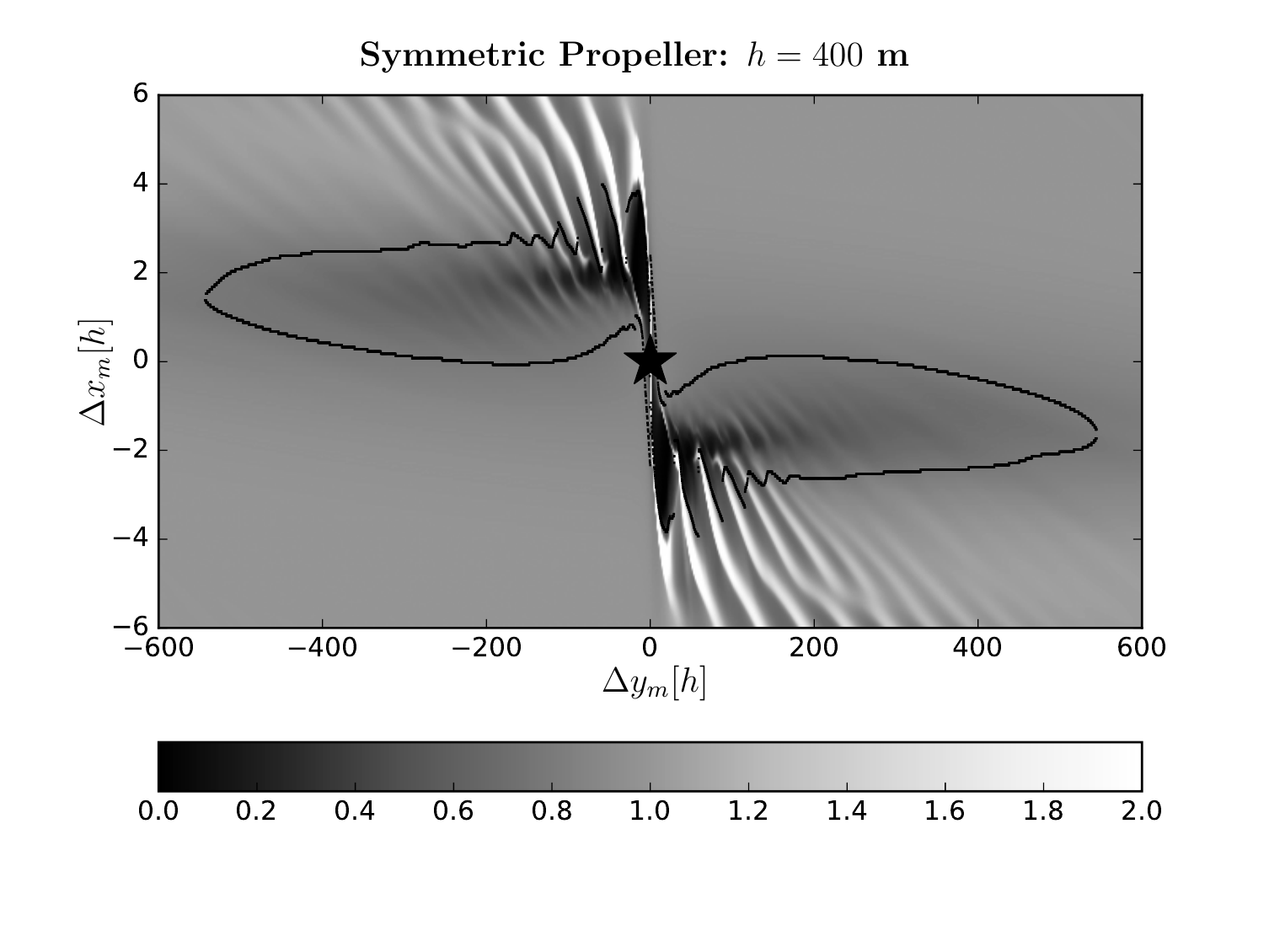}
\includegraphics[width=0.49\textwidth]{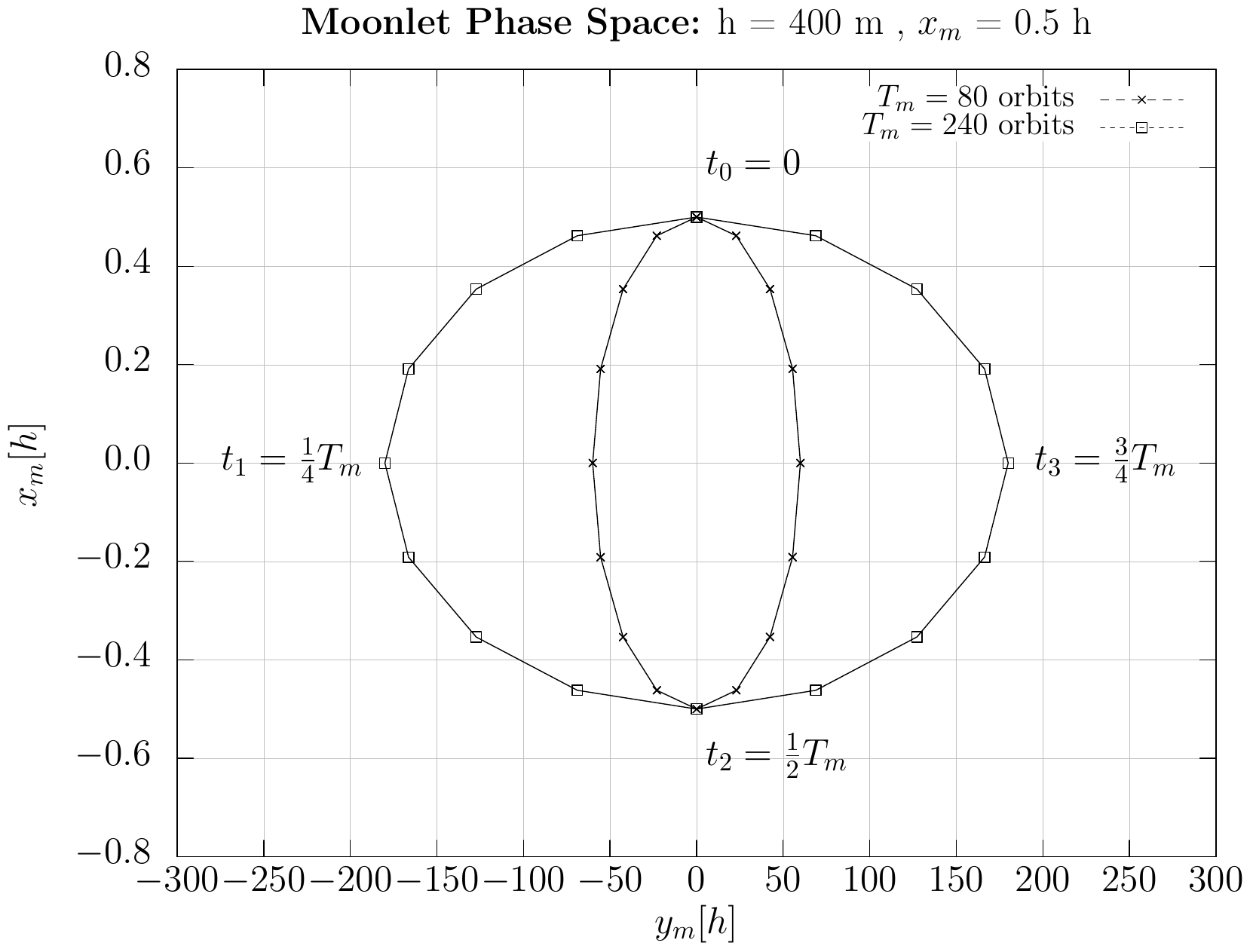}
\includegraphics[width=0.49\textwidth]{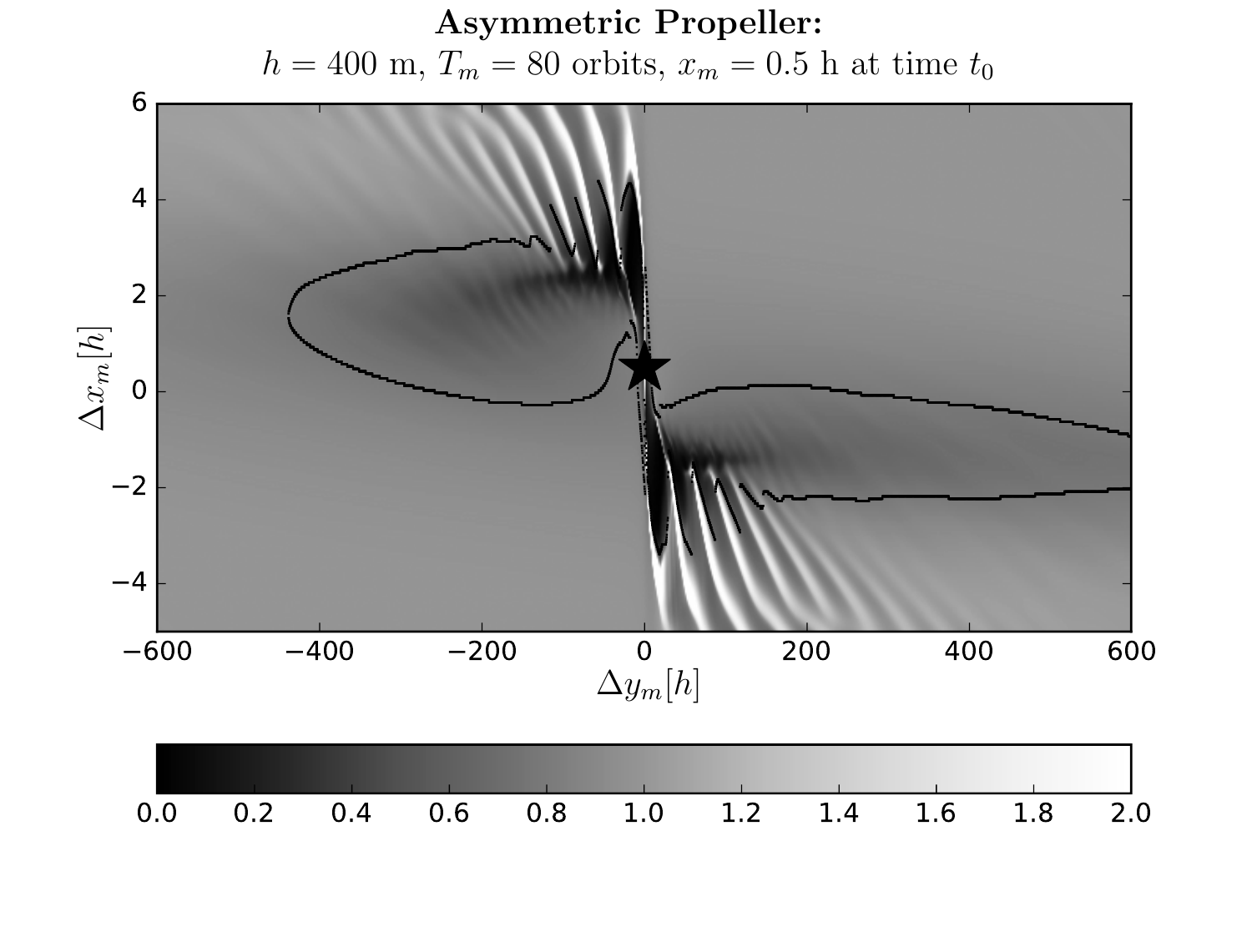}
\includegraphics[width=0.49\textwidth]{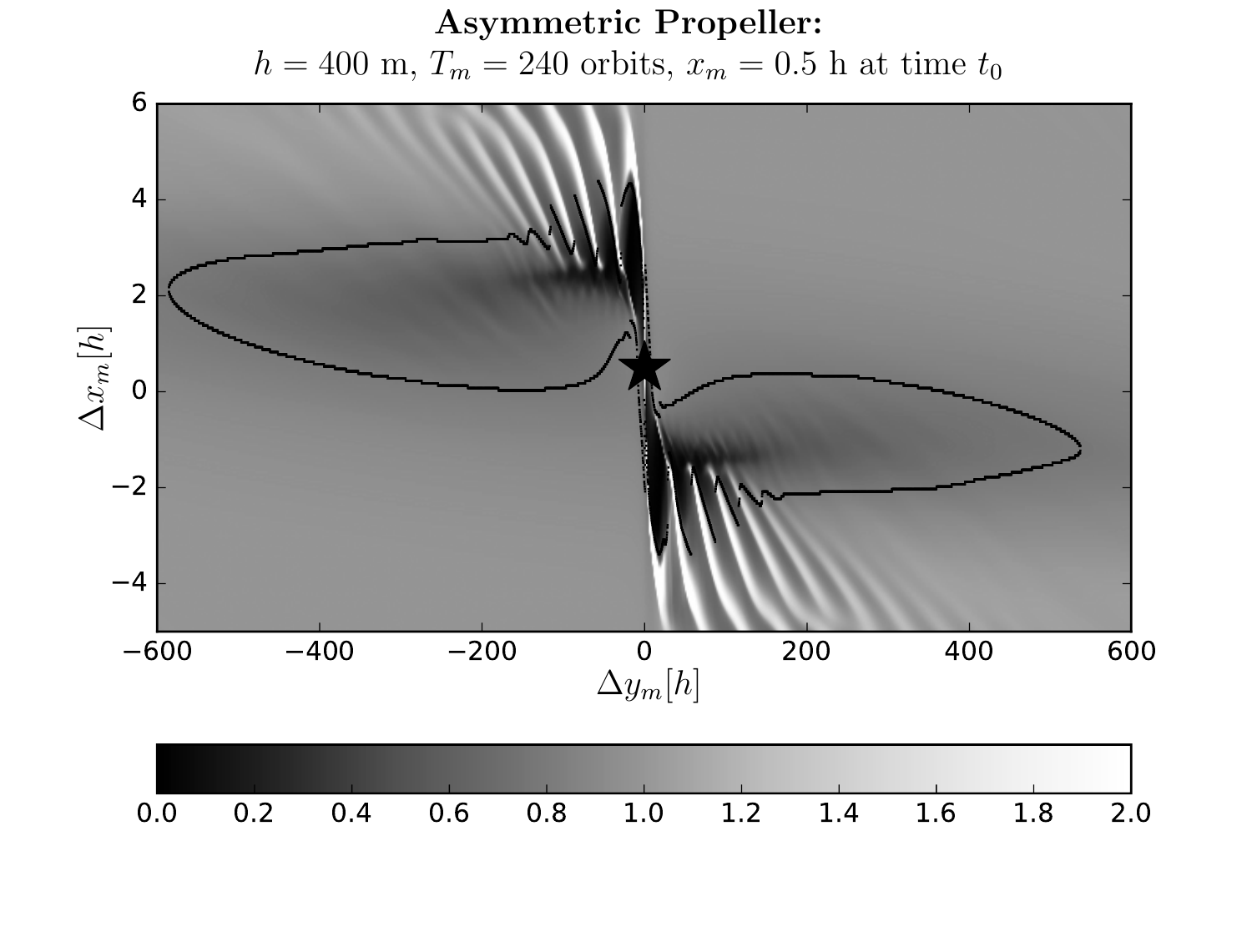}
\caption{\footnotesize Comparison of a symmetric propeller(top left) and two 
examples for an asymmetric propeller (bottom panels). The two asymmetric 
examples show the propeller for a librating moonlet at the same libration 
phase, but for different libration periods (see top right panel). 
In the bottom left panel the moonlet is librating with $T_{m} = T_{gap}$, 
while in the right one the period is $T_{m} = 3 \times T_{gap}$. In all 
panels, the origin of the coordinate system is centered to the mean orbital 
position of the moonlet. The moonlet's actual position is given by the 
star symbol. In all cases the moonlet's Hill radius is $h=\SI{400}{\meter}$ 
and its libration amplitude was set to $x_{m,0}= \SI{0.5}{\hill}$. 
The black lines denote the density profile at 80\% gap relaxation.}
\label{fig:density_plots}
\end{figure*}

The typical imprint of an unperturbed propeller structure, where the moonlet 
(represented by the star symbol) is not librating, is shown in the top panel 
of Figure \ref{fig:density_plots}. A clear symmetry can be seen. The grey scale 
color code illustrates the surface mass density, where darker color corresponds 
to regions of depleted density, while brighter color represents denser regions. 
The $x$ and $y$ directions in the presented plots denote the radial and 
azimuthal distances to the mean orbital position of the moonlet. For a better 
visualization, the presented area around the propeller is reduced radially and 
azimuthally to $\Delta x = \pm \SI{6}{\hill}$ and 
$\Delta y = \pm \SI{600}{\hill}$. Note, that the moonlet's mean orbital motion 
is from left to right. 

Fresh inflowing material (moving from left to right) passing by the by the 
moonlet from the inside at $\Delta x < 0$, are the more deflected to the 
moonlet the closer they pass it radially. The deflection for the particles 
passing by outside at $\Delta x > 0$ the moonlet moving from right to left 
occur vice versa. For larger radial distances, this deflection decreases as
$\left|\Delta x\right|^{-2}$. 
This induces a coherent motion of the ring material and thus results in the 
formation of wakes. The location of the moonlet within the simulation grid,
which is centered to its mean orbital location, is illustrated by the star 
symbol. Further, the black line marks the propeller surface mass density 
level at 80\% of gap closing ($\Sigma/\Sigma_0=0.8$). At this level of 
surface mass density, a clear symmetry can be seen when comparing the inner 
and outer gap region, where the gap lengths are about \SI{540}{\hill} and 
the maximum gap widths are about \SI{2}{\hill}, respectively.

This symmetry is broken in the bottom panels of Figure \ref{fig:density_plots}, 
where snapshots of the induced propeller structures for a librating moonlet are 
given. The azimuthal and radial scales are chosen exactly in the same way as 
for the unperturbed case. The level of $\Sigma/\Sigma_0 = 0.8$ is emphasized 
by the black line, where a clear difference in the gap lengths and even in the 
gap widths is obvious. The two bottom panels in Figure \ref{fig:density_plots} 
show the propeller structure at the same libration phase of the moonlet but for 
different libration periods. In the left panel the moonlet's libration period 
has the value $T_m=\SI{80}{\orbits}$, while its value is 
$T_m = \SI{240}{\orbits}$ in the right one. 
The iso-density level $\Sigma/\Sigma_0=0.8$ level is suitable to 
illustrate the asymmetry: The bottom left panel (shorter $T_m$) shows a 
length ratio between the outer and inner gap of 
$\SI{450}{\hill}/\SI{650}{\hill} \approx 0.7$ and in the other case 
(bottom right panel, $T_m = 3 \times T_{gap}$ this ratio takes 
$\SI{600}{\hill}/\SI{550}{\hill} \approx 1.09$, which makes the propeller gaps
look more symmetric. For the gap width at $\Delta = \pm \SI{150}{\hill}$ the 
width ratio between the outer and inner gap has a value of 
$\SI{3}{\hill}/\SI{2}{\hill} = 1.5$ for the bottom left panel, while for the 
bottom right panel this ratio is $\SI{3}{\hill}/\SI{2.5}{\hill} = 1.2$.

For the used dimension of the simulation box 
(see Table \ref{tab:further_params}) a steady state is already reached after 
\SI{80}{\orbits}, respectively. This equals the time the fresh ring material 
at $\Delta x = \pm \SI{1}{\hill}$ needs to drift to the moonlet. For the 
moonlet used with Hill radius of $h=\SI{400}{\meter}$ the azimuthal extent 
is of similar size (about \SI{600}{\hill} at $\Sigma/\Sigma_0=0.8$ as seen 
in the top panel of Figure \ref{fig:density_plots}). In this case, the 
perturbation by the moonlet motion would take about \SI{63}{\orbits} to reach 
the end of the gaps. Therefore, the two bottom panels of Figure 
\ref{fig:density_plots} show the induced propeller structure for 
$T_m\approx T_{gap}$ (left) and for $T_m\approx 3 \times T_{gap}$ (right). 
It has to be noted, that the timescales involved depend on the mass (size) 
of the moonlet.

\begin{figure*}
\centering
\includegraphics[width=0.49\textwidth]{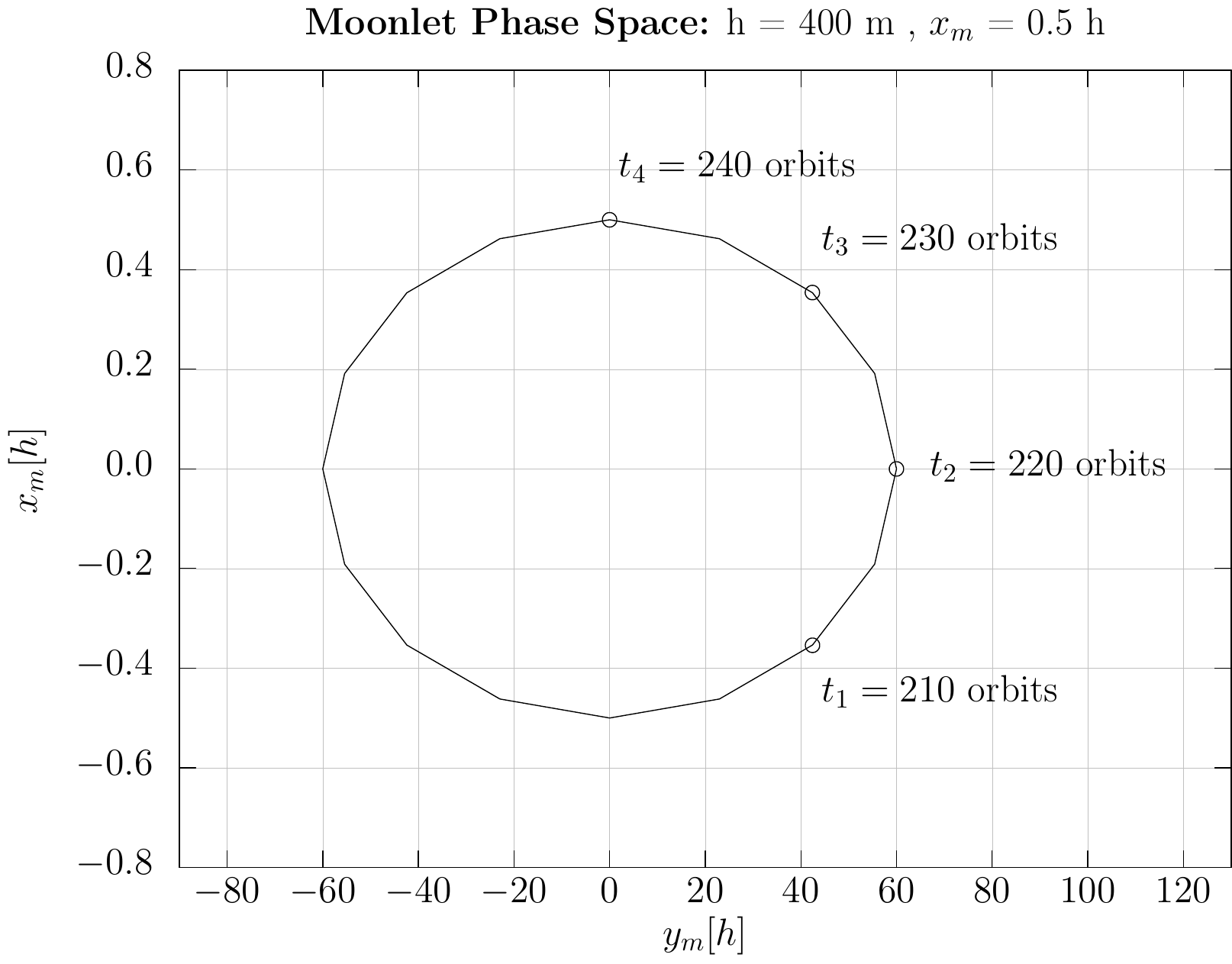}\\
\includegraphics[width=0.49\textwidth]{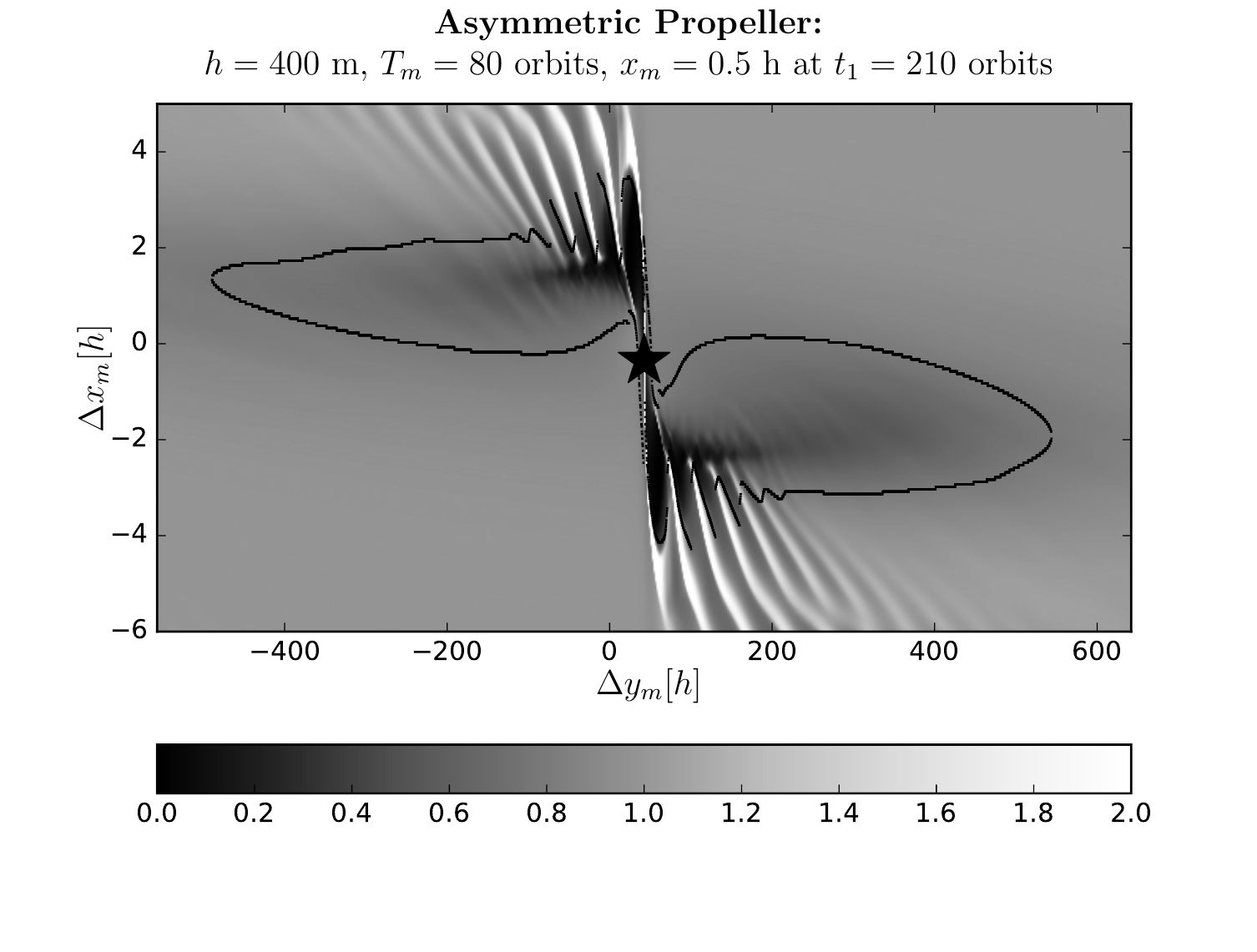}
\includegraphics[width=0.49\textwidth]{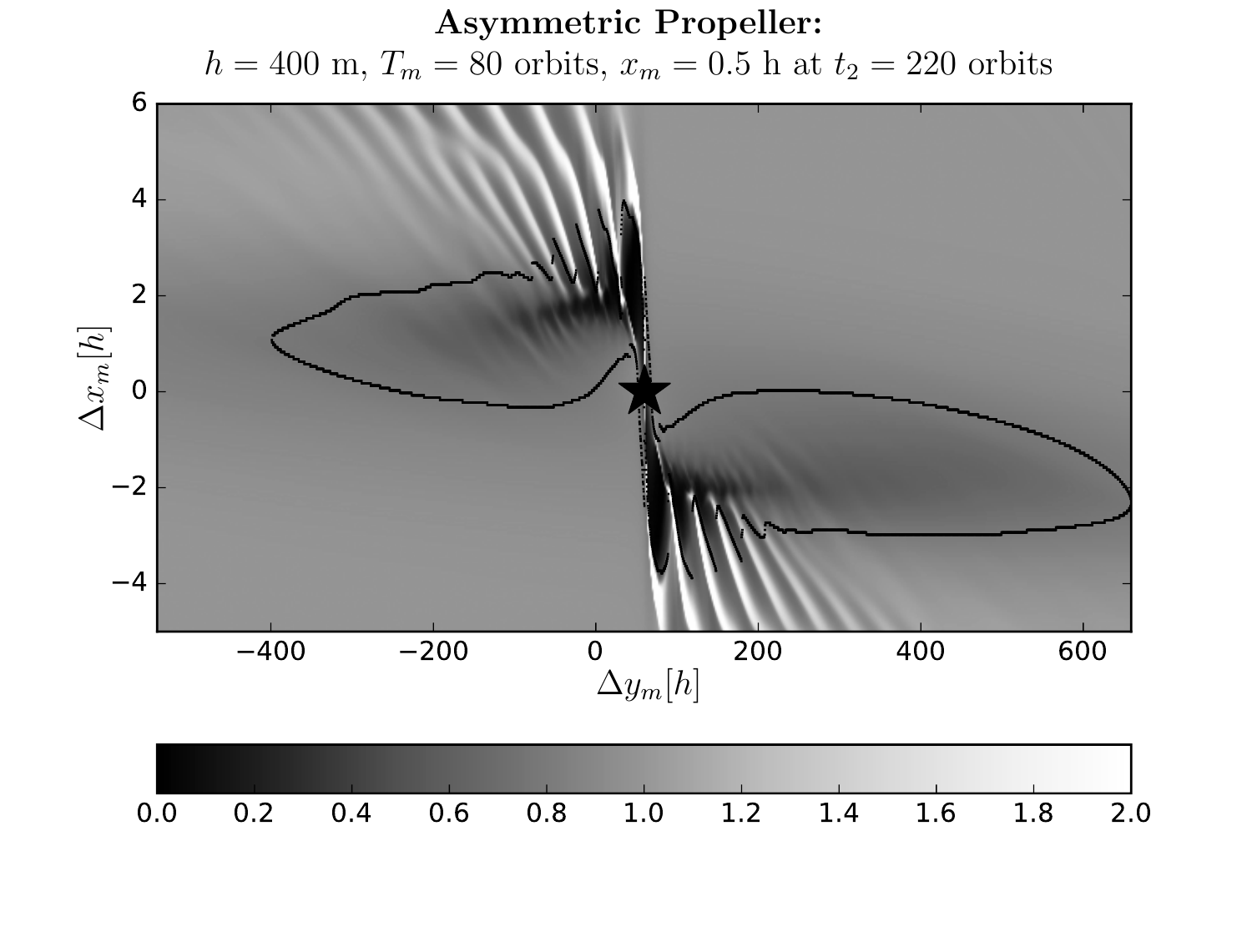}
\includegraphics[width=0.49\textwidth]{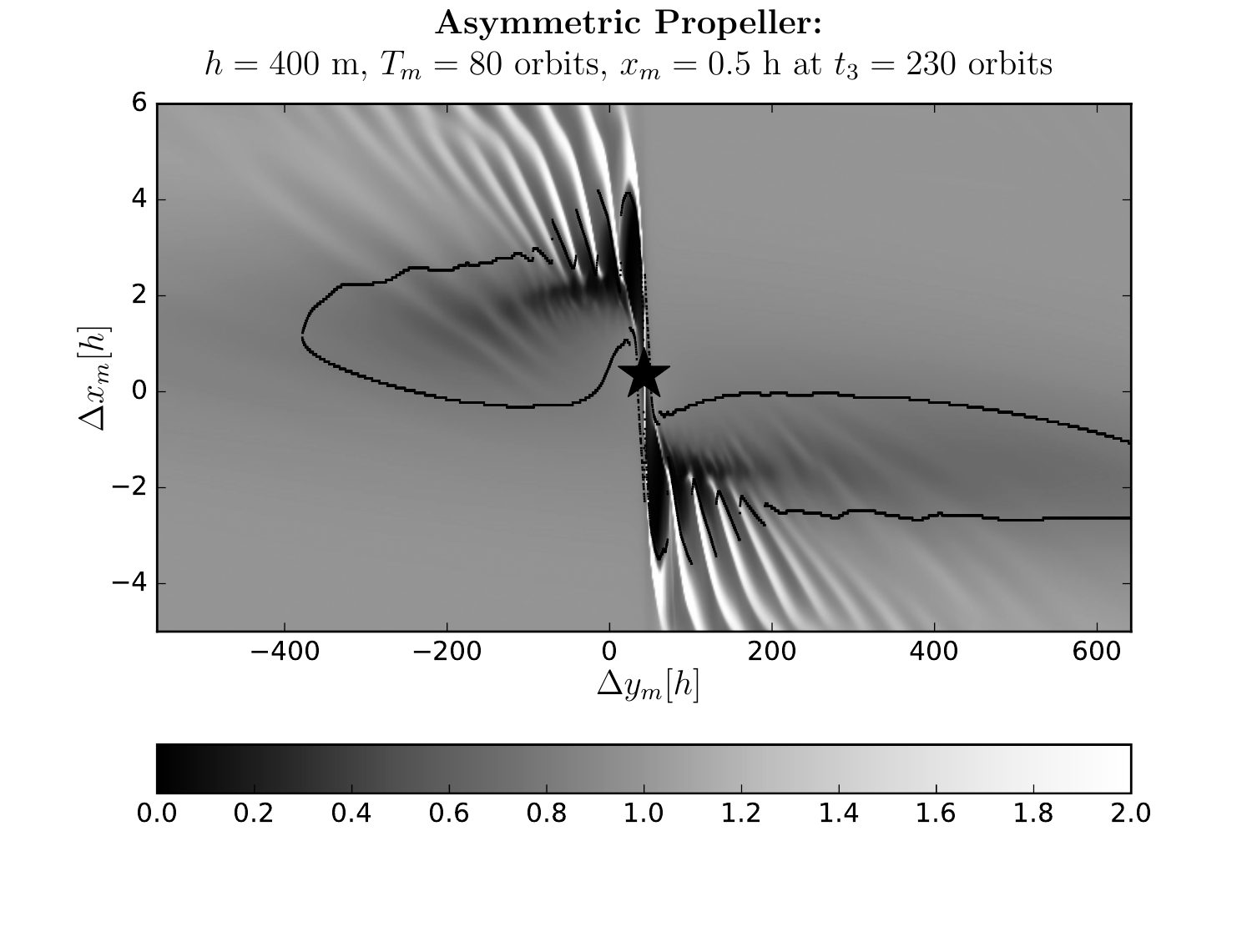}
\includegraphics[width=0.49\textwidth]{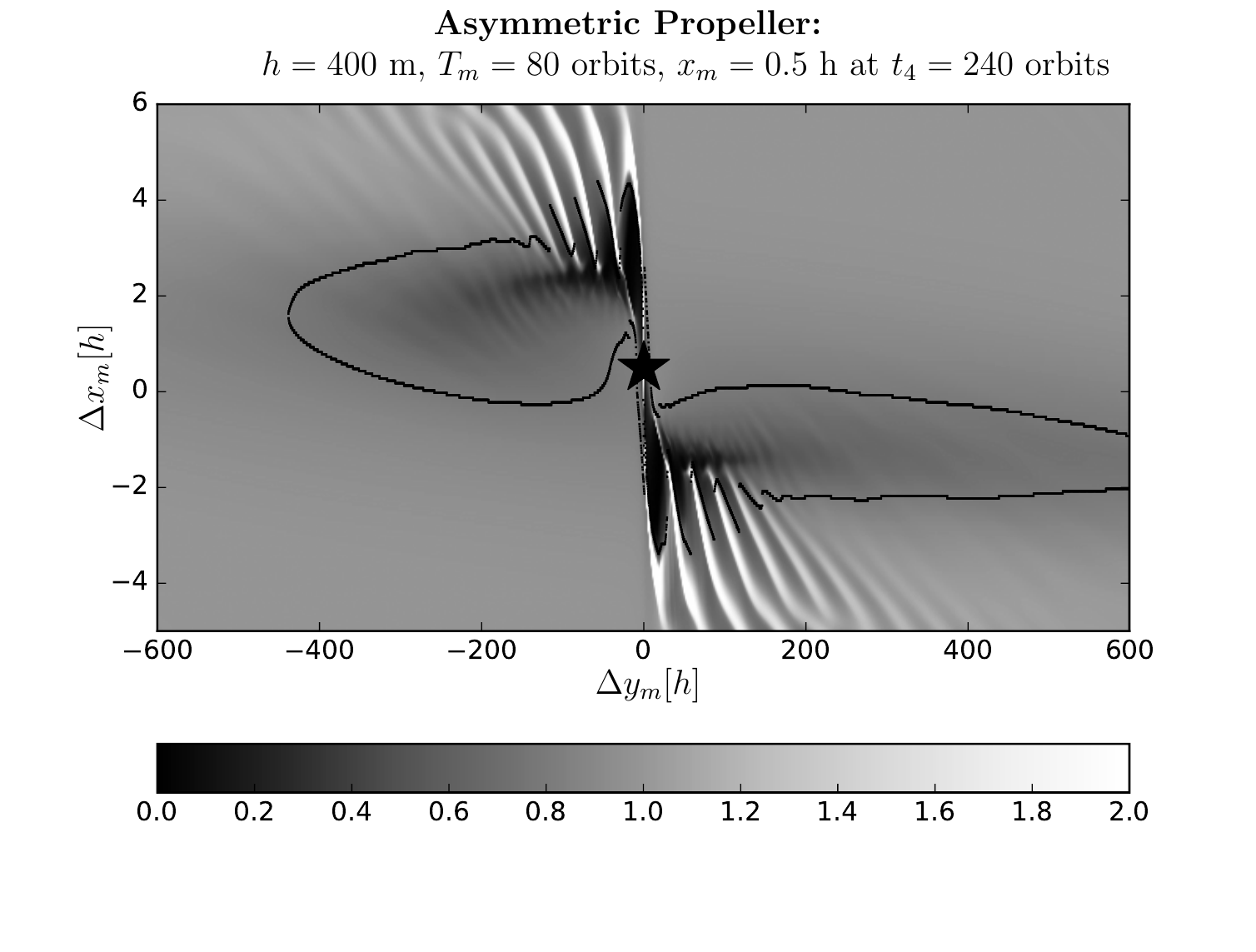}
\caption{\footnotesize Evolution of the propeller asymmetry for half a moonlet
period. Top panel: Phase space representation of the moonlet libration 
highlighting the four different libration phase angles shown in the center 
and bottom panels. The moonlet of size $h=\SI{400}{\meter}$ librates with 
a period of $T_m=\SI{80}{\orbits}$ and amplitude of $x_{m,0}=\SI{0.5}{\hill}$.
The snapshots (center and bottom panels) show the moonlet (star symbol) at 
different libration phases at times $t_1=\SI{210}{\orbits}$ (middle left), 
$t_2=\SI{220}{\orbits}$ (middle right), $t_3=\SI{230}{\orbits}$ 
(bottom left), and $t_4=\SI{240}{\orbits}$ (bottom right).} 
\label{fig:density_plots_time_series}
\end{figure*}

Due to the libration of the embedded moonlet, the shape of the propeller is
changing all the time. Thus, at different phases of the moonlet's libration
the propeller structure looks different resulting in different gap lengths 
and gap depths, as presented in Figure \ref{fig:density_plots_time_series}. 
Here, the four center and bottom panels show the propeller shape for half a 
libration period at times 
$t = \numlist[list-final-separator = { , }]{210;220;230} \; \text{and} \;
\SI{240}{\orbits}$, while the top panel illustrates the actual position of 
the moonlet at those times. Note, that the moonlet libration is counter 
clockwise. The propeller shape for the other half of the libration period 
would look similar but mirrored due to the Kepler shear. The snapshots at 
\SI{220}{\orbits} and \SI{240}{\orbits} show the moonlet at 
$(x_m(t=\SI{220}{\orbits})=\SI{0}{\hill},y_m(t=\SI{220}{\orbits})=\SI{60}{\hill})$ and
$(x_m(t=\SI{240}{\orbits})=\SI{0.5}{\hill},y_m(t=\SI{240}{\orbits})=\SI{0}{\hill})$. 
The strongest asymmetry of the propeller can be observed for the moonlet at 
its radial amplitude (see snapshot at $t=\SI{220}{\orbits}$ in Figure 
\ref{fig:density_plots_time_series}).

A more detailed analysis of the gap structures will be presented in the
following sections.

\subsection{Radial Gap Profile} \label{sec:gap_profile}
Figure \ref{fig:density_plots_time_series} illustrates the evolution of the
asymmetry of a propeller caused by a librating moonlet. The grey-level density
representations correspond to certain times 
$t=(\num{210},\num{220},\num{230},\num{240})$ orbits, which represent four
different phases (phase angles) of the moonlet libration.

We will analyze and compare the radial gap profiles of a symmetric and 
asymmetric propeller at different azimuthal distances to the moonlet which 
demonstrate the changes in the gap structures due to the moonlet motion.
The minimum of the surface mass density defines the gap depth and its radial 
position gives the gap position. At the fixed surface density level
$\Sigma/\Sigma_0 = 0.8$ we estimate the gap width, permitting the comparison 
of the inner and outer gap in this way.

\begin{figure*}
\centering
\includegraphics[width=0.49\textwidth]{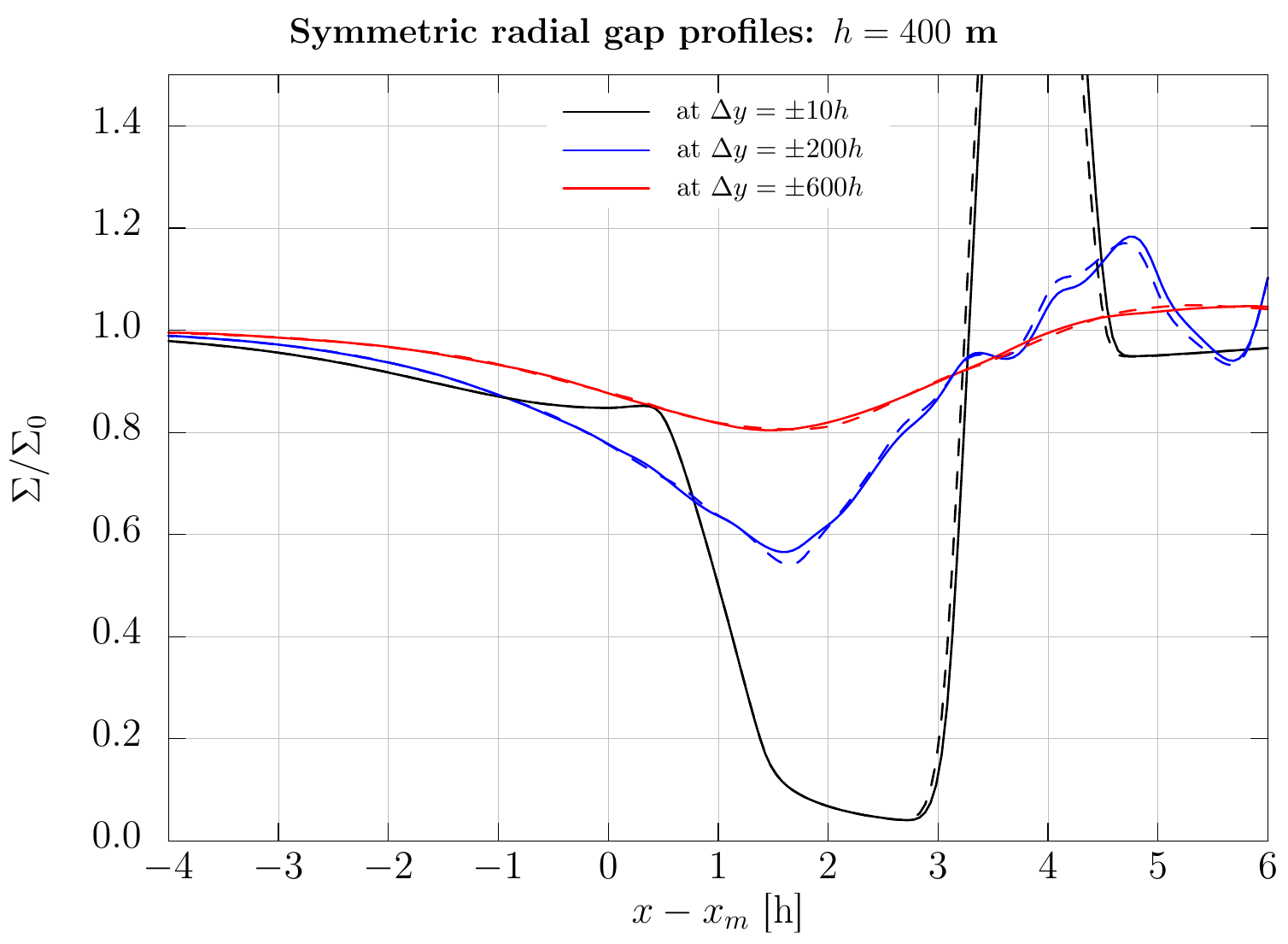}\\
\includegraphics[width=0.49\textwidth]{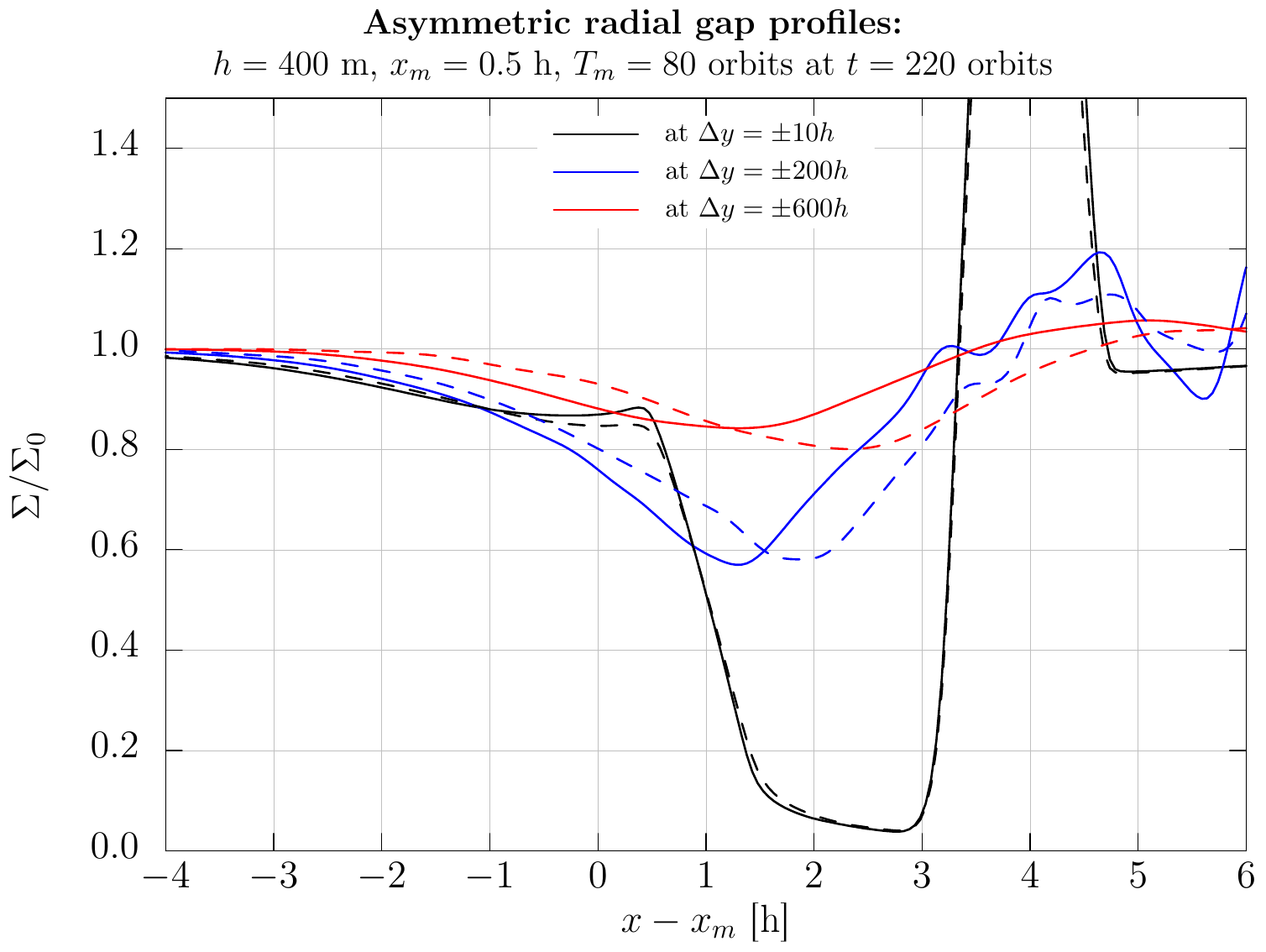}
\includegraphics[width=0.49\textwidth]{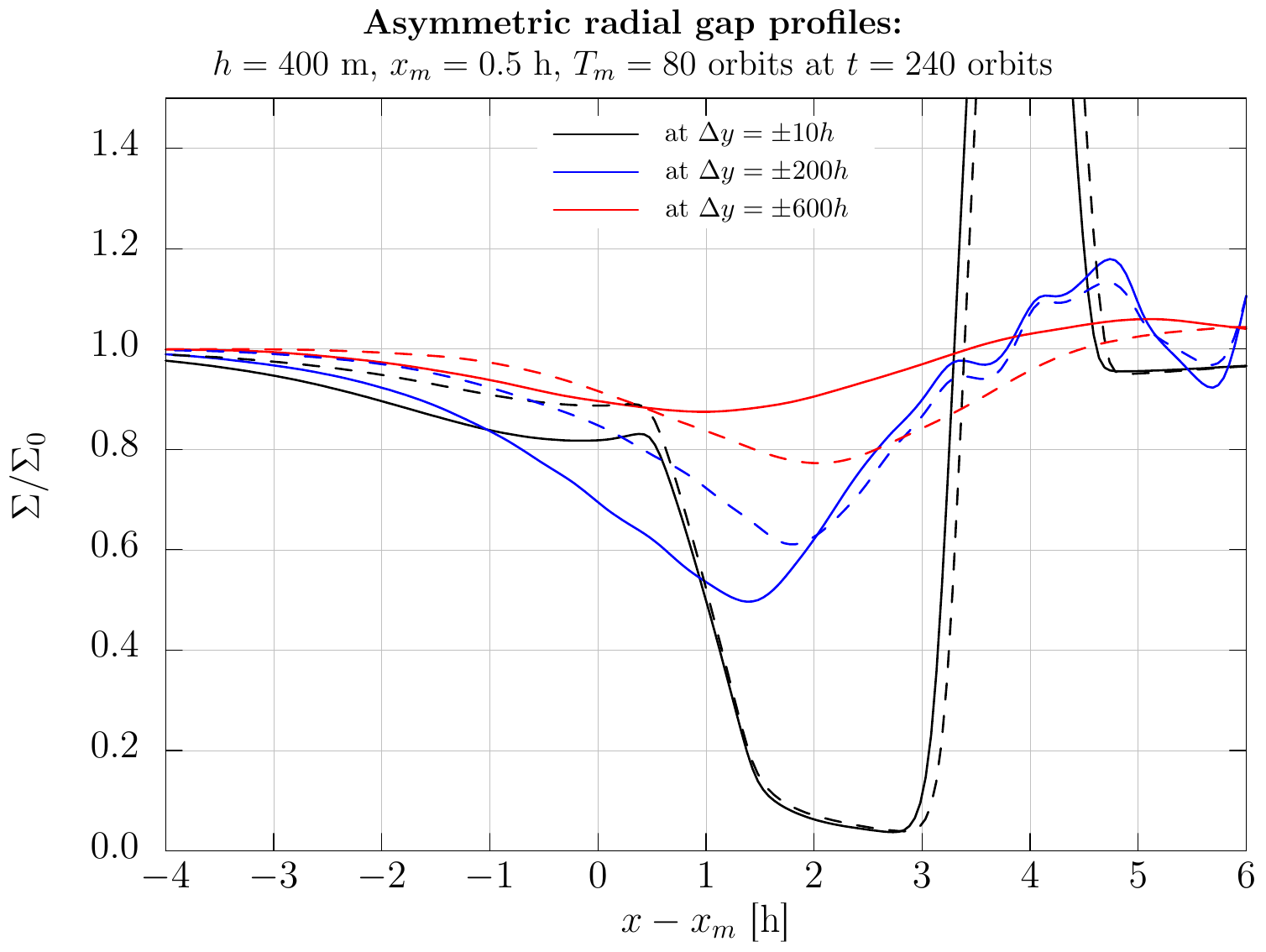}
\caption{\footnotesize Radial gap profiles for a symmetric (Top) 
and asymmetric propeller structure (Bottom Left and Right) at different 
azimuthal distances to the moonlet of $\Delta y=\pm$\SI{10}{\hill} (black), 
$\pm$ \SI{200}{\hill} (blue), $\pm$\SI{600}{\hill} (red). The dashed and solid 
lines denote the outer and inner gap profile. The inner gap's profile has been
mirrored at the moonlet position for better comparison. 
For the asymmetric propeller (Bottom panels) the radial gap profile is given 
at two different libration phases of the moonlet. Here, the moonlet size has 
been set to $h=\SI{400}{\meter}$ and the libration period and amplitude have 
been set to \SI{80}{\orbits} and \SI{0.5}{\hill}. For better comparison, the 
moonlet's current position has been subtracted of in the case of the 
asymmetric propeller.}
\label{fig:RadialProfile}
\end{figure*}

First, we reduce the noise of the simulation data -- especially in the wake 
region close to the moonlet -- by performing moving box averaging for the 
radial gap profiles, where the radial box length is set close to the wake 
wavelength of about $\Delta x = \SI{0.5}{\hill}$. The resulting smoothed 
radial density profiles of the inner and outer gap at azimuthal distances 
to the moonlet of 
$\Delta y=\left|y-y_m\right| = \numlist[list-final-separator = {\text{ and
}}]{10;200;600} \, \text{h}$ are shown in Figure \ref{fig:RadialProfile}. 
Note, that for the bottom panels in Figure \ref{fig:RadialProfile} we use a
moonlet-centered coordinate system ($\Delta x = x-x_m(t)$ and
$\Delta y = y-y_m(t)$) where we subtract of the moonlet's actual position 
for better comparison to the unperturbed moonlet profiles. The solid and 
dashed curves denote the inner and outer gap profiles. Note, that we 
mirrored the inner gap's profile at the origin to compare both gap profiles 
in a better way.

The top panel of Figure \ref{fig:RadialProfile} shows the radial profile for 
the symmetric propeller. Here, the gap locations of the inner and outer gap 
match at all azimuthal distances. Further, the gap relaxes azimuthally as
theoretically predicted, i.e. the gap-depth decreases for increasing 
azimuthal distance to the moonlet. 

In the bottom panels the asymmetric gap profiles are illustrated at the two
snapshots $T=\SI{220}{\orbits}$ and $T=\SI{240}{\orbits}$ from Figure 
\ref{fig:density_plots_time_series}. 
Here, the moonlet is presented at its azimuthal (left) and radial (right)
libration amplitude. While in the left panel the gap profiles look much alike, 
but are shifted radially with increasing azimuth, their depths and widths 
differ significantly in the right panel. Note, how the beginning of the gaps 
(compare $\Delta y=\SI{10}{\hill}$) stays unaffected by the moonlet libration 
and follows the moonlet motion instantaneously in both panels.

\subsubsection{Gap Positions}
\begin{figure*}
\centering
\includegraphics[width=0.49\textwidth]{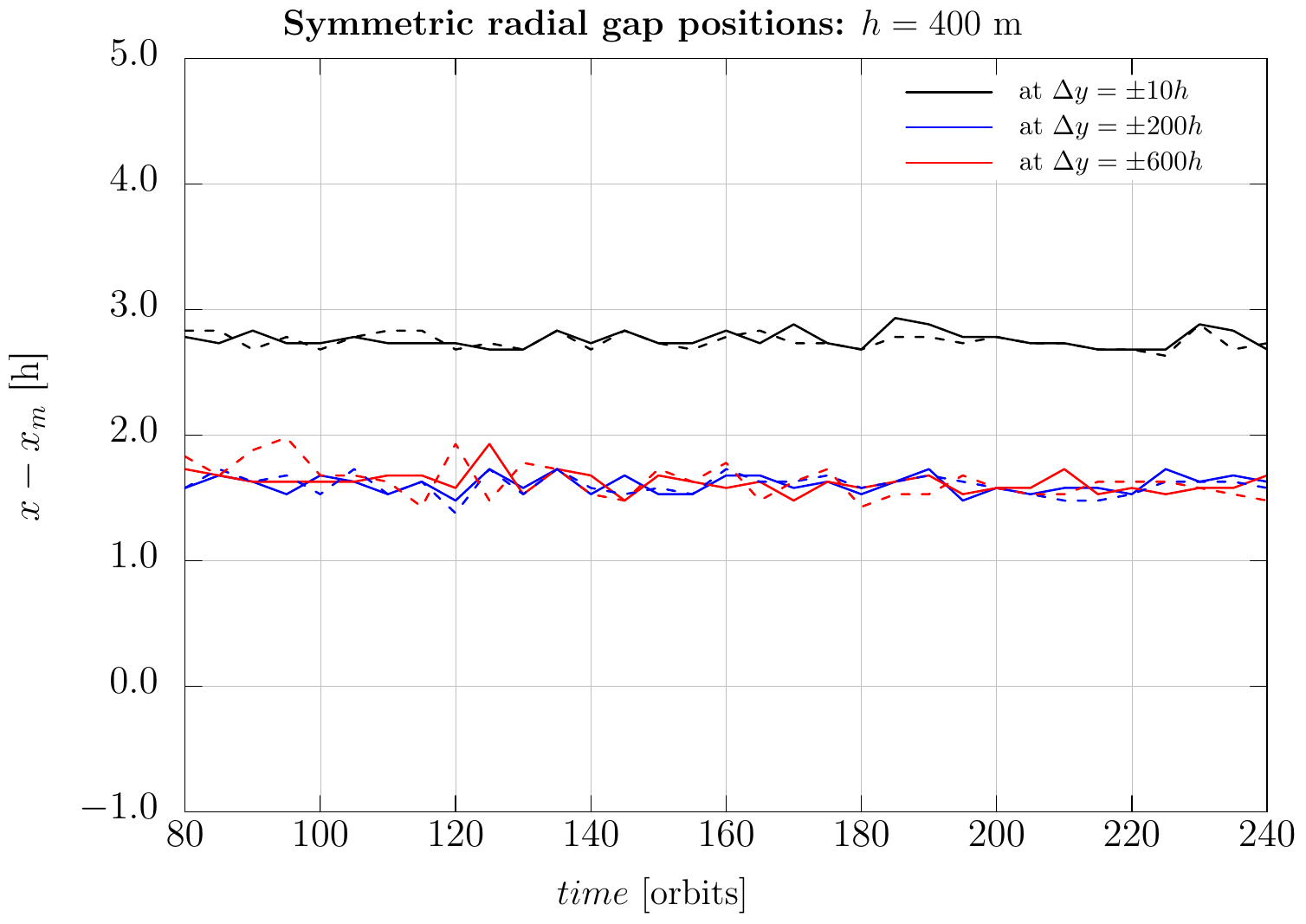}
\includegraphics[width=0.49\textwidth]{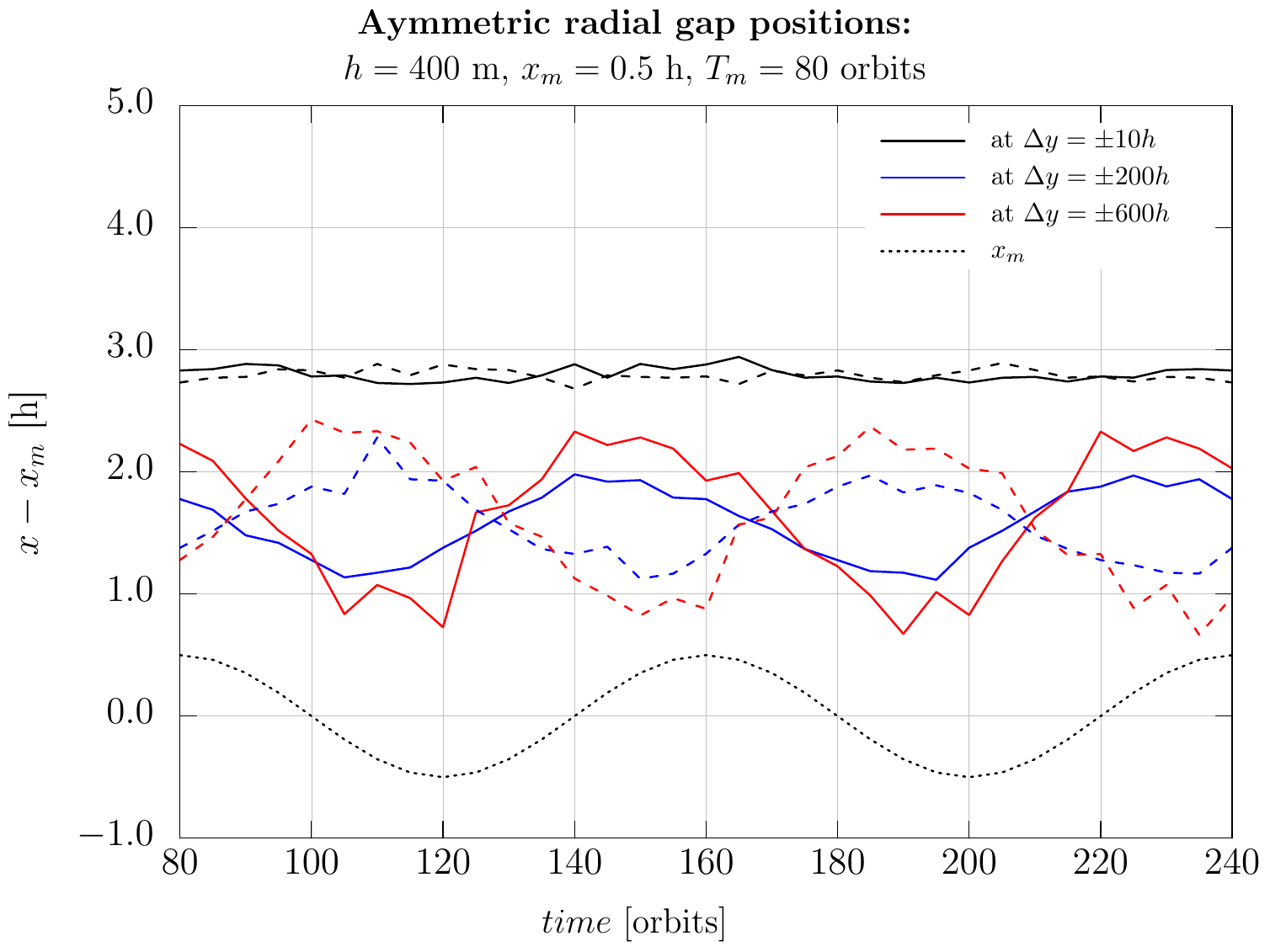}
\caption{\footnotesize 
Evolution of the radial gap minima location 
for the symmetric (left) and asymmetric (right) propeller at azimuthal 
distances to the moonlet of $\Delta y=\pm$\SI{10}{\hill} (black), $\pm$ 
\SI{200}{\hill} (blue) and $\pm$\SI{600}{\hill} (red). The coordinate system 
is moonlet-centered, meaning, that the current moonlet posotion 
$\Delta x = x-x_m(t)$ and $\Delta y = y-y_m(t)$ has been subtracted of. 
The dashed and solid lines denote the outer and inner gap profile. 
The moonlet's radial position in the right panel is shown by the black dotted 
line. An increasing radial amplitude of the changing gap minima locations can 
be seen for larger azimuths. For the simulation the moonlet has the Hill radius 
$h=\SI{400}{\meter}$ and was librating with a period and amplitude of 
\SI{80}{\orbits} and \SI{0.5}{\hill}.}
\label{fig:RadialProfileEvo}
\end{figure*}

The changes in the gap minima locations caused by the moonlet libration can be 
better illustrated by considering their time-evolution for a fixed azimuth, as
shown in Figure \ref{fig:RadialProfileEvo}, where the evolution of the gap 
minima locations at 
$\Delta y=\left|y-y_m\right| = \numlist[list-final-separator = {\text{ and
}}]{10;200;600} \, \text{h}$ is presented.

For the symmetric propeller the radial gap positions for different azimuths
lay on top of each other, where the mean radial gap position $<x_{gap}> \approx
\pm\SI{2.83}{\hill}$ (at $\Delta y = \pm \SI{10}{\hill}$) decreases to 
$<x_{gap}> \approx \SI{1.6}{\hill}$ (for $\Delta y = \pm\num{200}$ and 
$\pm\SI{600}{\hill}$) for larger azimuthal distances.

Figure \ref{fig:RadialProfileEvo} shows the time-evolution of the gap minima
location for the symmetric (left panel) and the asymmetric (right panel) 
propeller. For the asymmetric case, it is visible, how the locations of the 
gap minima clearly follow the moonlet libration (black dotted line). 
Further, the radial amplitude of the gap location slightly increases for 
larger azimuthal distance to the moonlet. For the symmetric case, the radial 
gap locations stay constant over time.

While the ring material in the close vicinity to the moonlet almost immediately 
feels the change in the moonlet's motion (compare the the black solid and dashed
lines in Figure \ref{fig:RadialProfileEvo}), the perturbation by the motion of 
the moonlet first needs to be transported through the ring environment to larger 
azimuthal distances 

\begin{equation}
  \Delta y = -\frac{3}{2} \Omega_m \, t \, \Delta x
\end{equation}

set by the Kepler shear. Thus, when the moonlet reaches its radial libration 
amplitude of \SI{0.5}{\hill} at $T=\SI{80}{\orbits}$ the gap minima at 
$\Delta y=\SI{200}{\hill}$ and $\Delta y=\SI{600}{\hill}$ are following the 
moonlet motion after about \SI{26}{\orbits} and \SI{80}{\orbits}, respectively. 
This agrees with the delay found in Figure \ref{fig:RadialProfileEvo}.

\subsubsection{Radial Gap Separation} \label{sec:gap_separation}
\begin{figure*}
\centering
\includegraphics[width=0.49\textwidth]{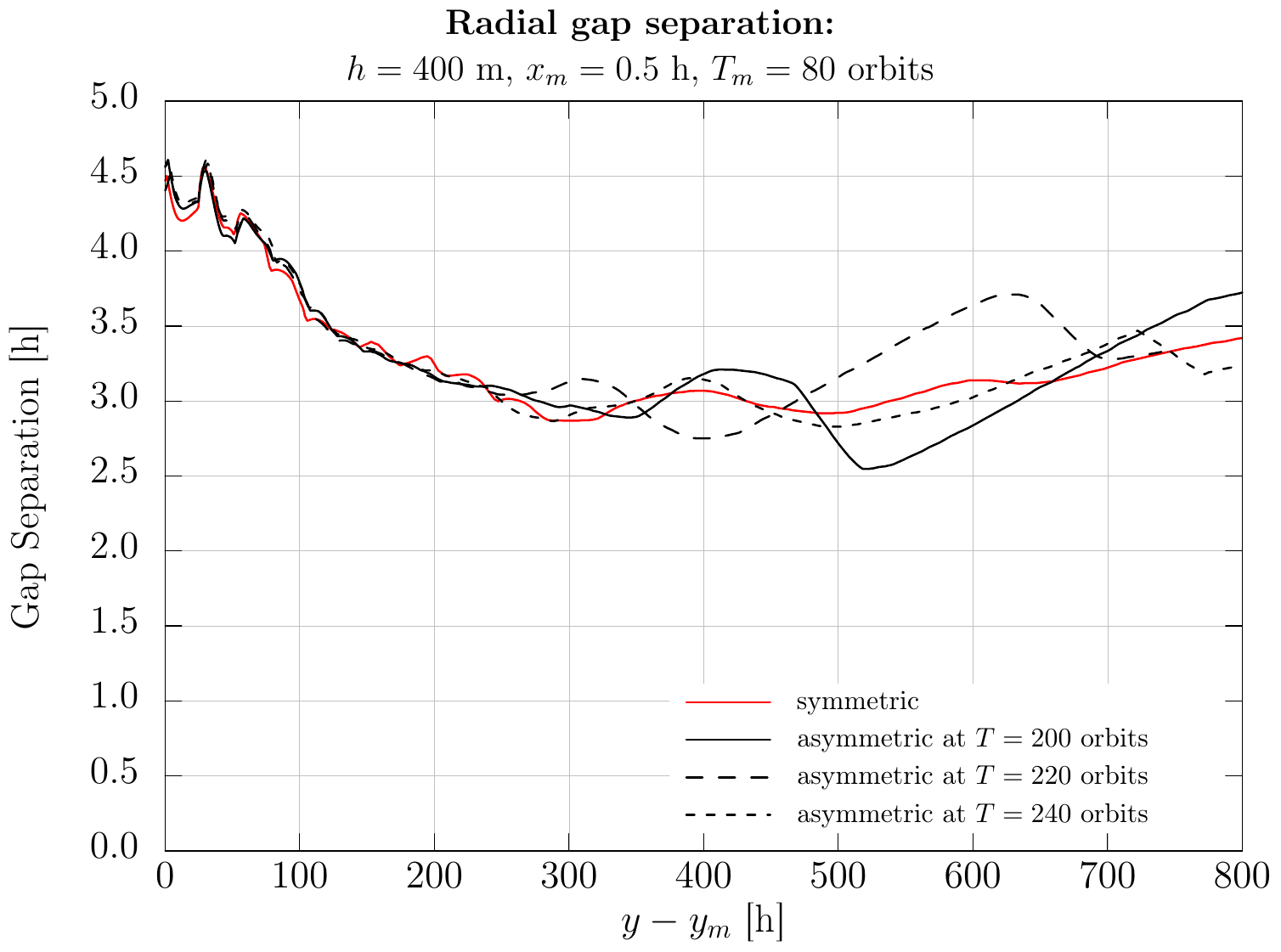}
\includegraphics[width=0.49\textwidth]{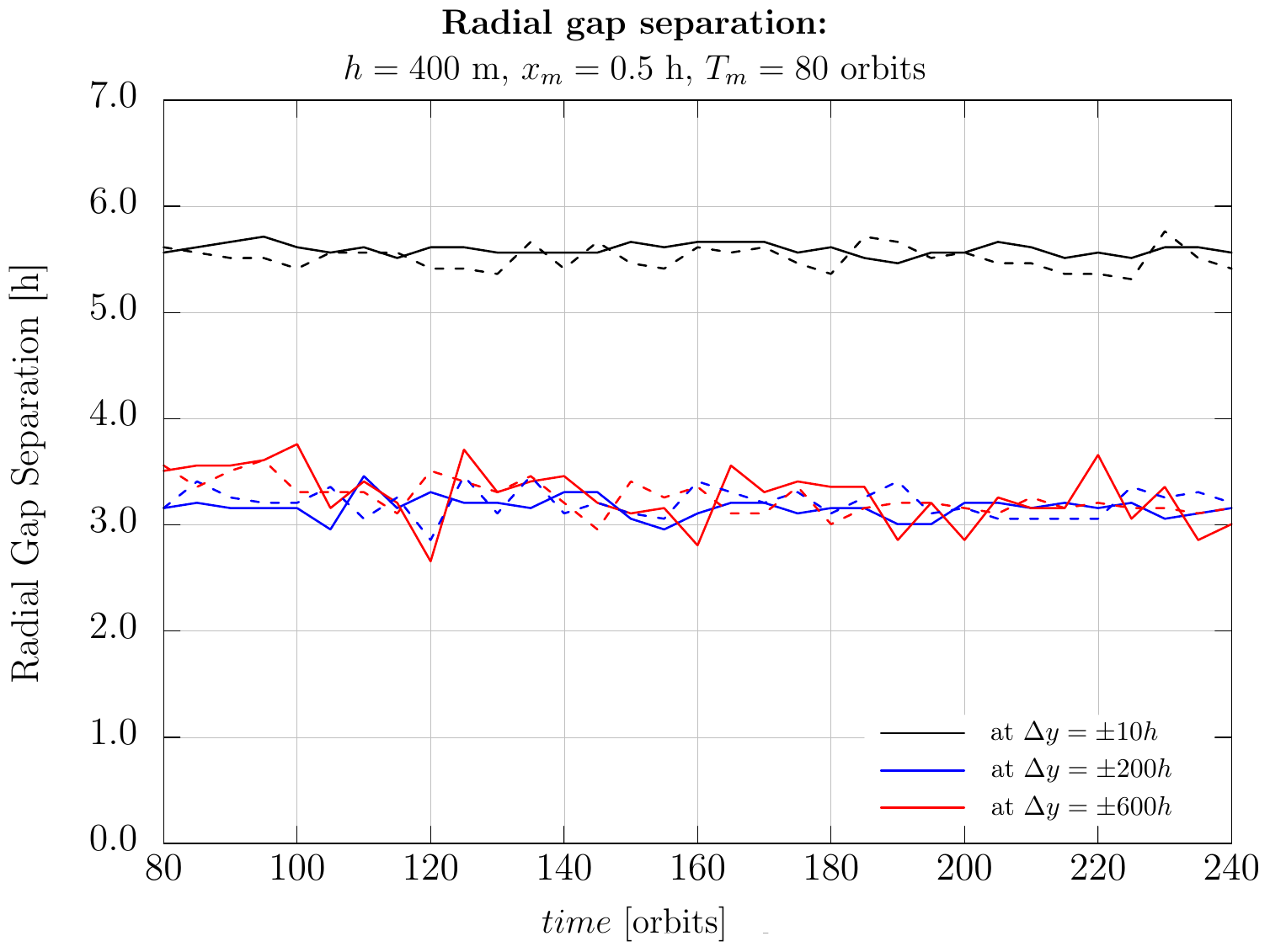}
\caption{\footnotesize
Gap Separation for a symmetric and asymmetric propeller structure for a moonlet
with $h=\SI{400}{\hill}$ Hill radius. For the asymmetric propeller structure 
the moonlet was librating with a period of \SI{80}{\orbits} and an amplitude of
\SI{0.5}{\hill}. For all realizations, a moonlet-centered reference frame has
been used. 
Left: Radial gap separation along the azimuth. Red and black solid lines denote 
the symmetric and asymmetric propeller structure. The asymmetric propeller
structure is presented at different snapshots at $T=\num{200}, \num{220}$ and
\SI{240}{\orbits}. Right: Radial gap separation for a symmetric (dashed) and 
asymmetric (solid) propeller structure at azimuthal distances to the moonlet of
$\Delta y=\pm\SI{10}{\hill}$ (black), $\Delta y=\pm\SI{200}{\hill}$ (blue) and
$\Delta y=\pm\SI{600}{\hill}$ (red).}
\label{fig:GapSeparation}
\end{figure*}

Although the radial locations of the gap minima change over time and can differ
by more than \SI{1}{\hill} (compare the solid and dashed red lines in Figure 
\ref{fig:RadialProfileEvo}), their separation stays almost constant over time
(see Figure \ref{fig:GapSeparation}).
The evolution of the radial gap separation is presented at azimuthal distances 
to the moonlet of 
$\Delta y=\left|y-y_m\right| = \numlist[list-final-separator = {\text{ and 
}}]{10;200;600}$ h, where the dashed and solid lines denote the symmetric and
asymmetric gap separation, respectively. 
With increasing azimuth the scattering of the radial gap separation becomes 
larger (about \SI{1}{\hill} at $\Delta y = \pm \SI{600}{\hill}$ and 
\SI{0.5}{\hill} at $\Delta y = \pm \SI{200}{\hill}$). 
This is illustrated in the left panel of Figure \ref{fig:GapSeparation}, where 
a comparison of the gap separation for the symmetric propeller structure 
(red solid line) along the azimuth is compared against the asymmetric one 
(black lines) for three different snapshots. 
Comparing all the curves, one recognizes, that up to a critical azimuth of about 
\SI{250}{\hill} the variations of the radial gap separation for the asymmetric
gaps are negligible and thus it almost matches the symmetric gap structure.

For larger azimuthal distances to the moonlet, the retardation effect gets more 
dominant resulting in larger variations of the radial gap separation along the 
azimuth with time. This allows to detect the perturbation by the moonlet motion 
which gets visible as a wavy structure along the azimuth. Further, comparing the 
different snapshots at times $T=\SI{200}{\orbits}$ and $T=\SI{220}{\orbits}$ 
the propagation of the perturbation along the azmiuth can be observed. 
On average, the gap separation still follows the azimuthal behavior of the 
unperturbed gap separation.

\subsubsection{Gap Width} \label{sec:gap_width}
The grey-level density plots in Figure \ref{fig:density_plots} and the 
radial profiles presented in Figure \ref{fig:RadialProfile} have 
demonstrated, that the gap width is influenced by the motion of the 
moonlet. We study the effect of the moonlet motion on the gap 
width by estimating the gap width at 80 \% gap closing 
($\Sigma/\Sigma_0=0.8$).

\begin{figure*}
\centering
\includegraphics[width=0.49\textwidth]{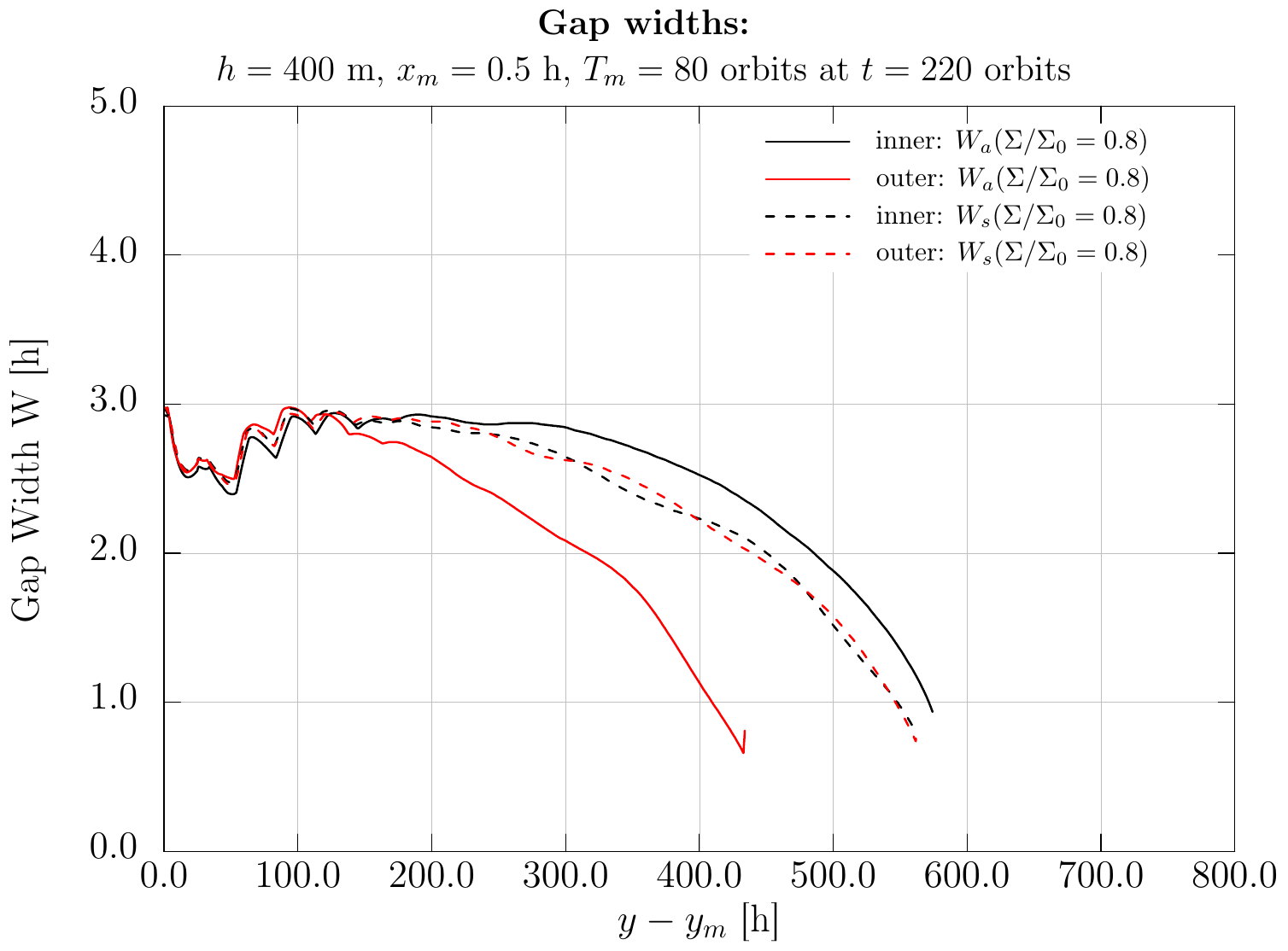}
\includegraphics[width=0.49\textwidth]{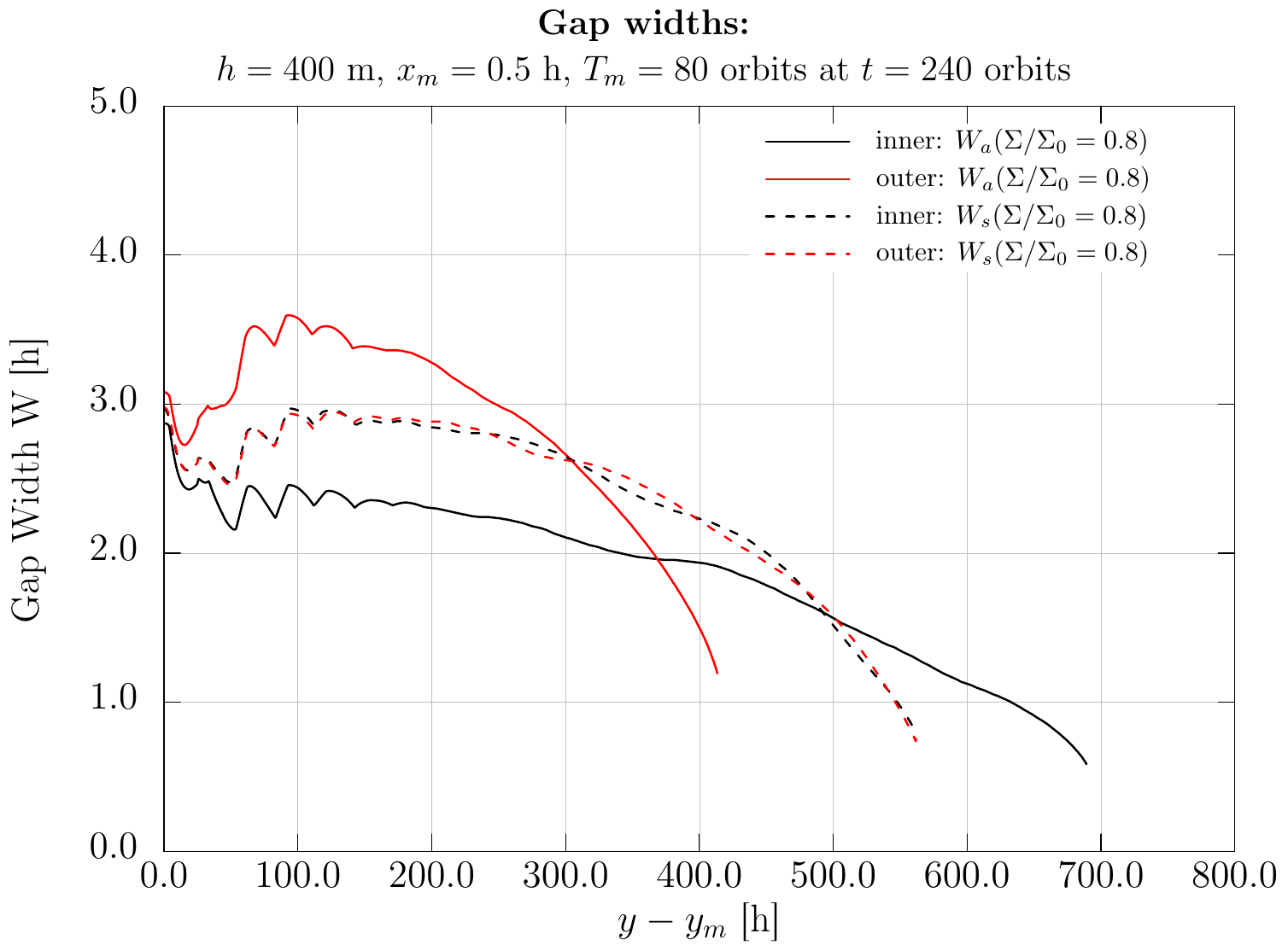}
\caption{\footnotesize Comparison of the gap width at $\Sigma/\Sigma_0
= 0.8$ for the symmetric (dashed) and asymmetric propeller (solid) at 
$T=\SI{220}{\orbits}$ (left) and $T=\SI{240}{\orbits}$ (right).  The widths of 
the inner and outer gap are given by the black and red colored lines. The
results have been obtained from simulations of a $h=\SI{400}{\meter}$ sized
moonlet, wich was librating with a radial amplitude and period of 
\SI{0.5}{\hill} and \SI{80}{\orbits} for the asymmetric case.}
\label{fig:GapWidth}
\end{figure*}

The resulting gap widths for the inner and outer gap are shown in Figure
\ref{fig:GapWidth}. A concave decrease for all curves can be seen, resulting 
from the azimuthal gap relaxation. A similar trend holds for the gap depths. 

In the panels of Figure \ref{fig:GapWidth} the dashed lines refer to
the symmetric propeller structure, whereas the solid lines represent the 
widths of the asymmetric gap structures. 
The red and black colors denote the outer and inner gap structure. In all 
representations the gap widths of the symmetric and asymmetric gaps differ 
significantly. The moonlet libration can be identified by the intersection 
points of the inner and outer gap width profiles (right panel, Figure
\ref{fig:GapWidth}), where the gap widths exchange their roles.

\begin{figure*}
\centering
\includegraphics[width=0.49\textwidth]{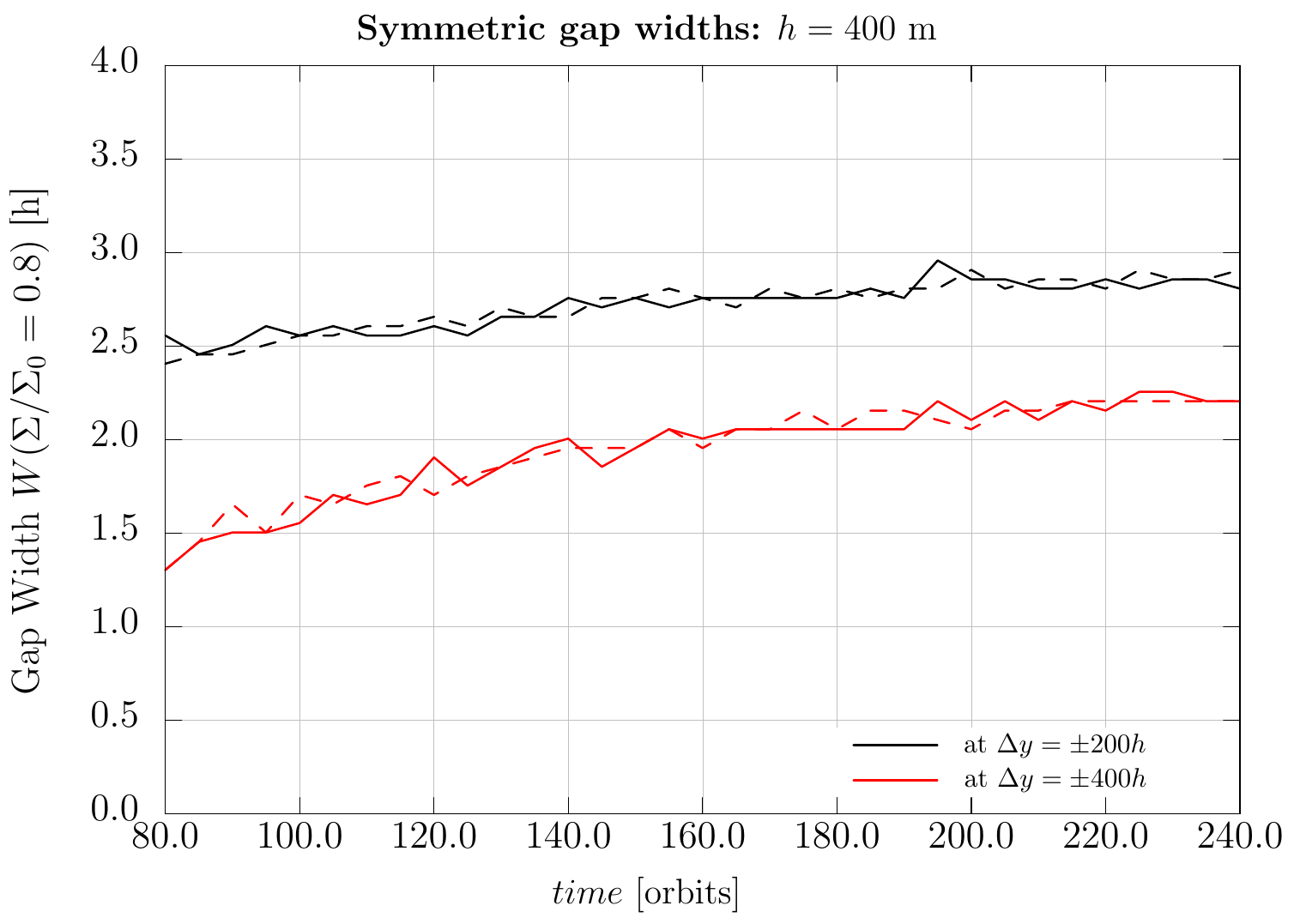}
\includegraphics[width=0.49\textwidth]{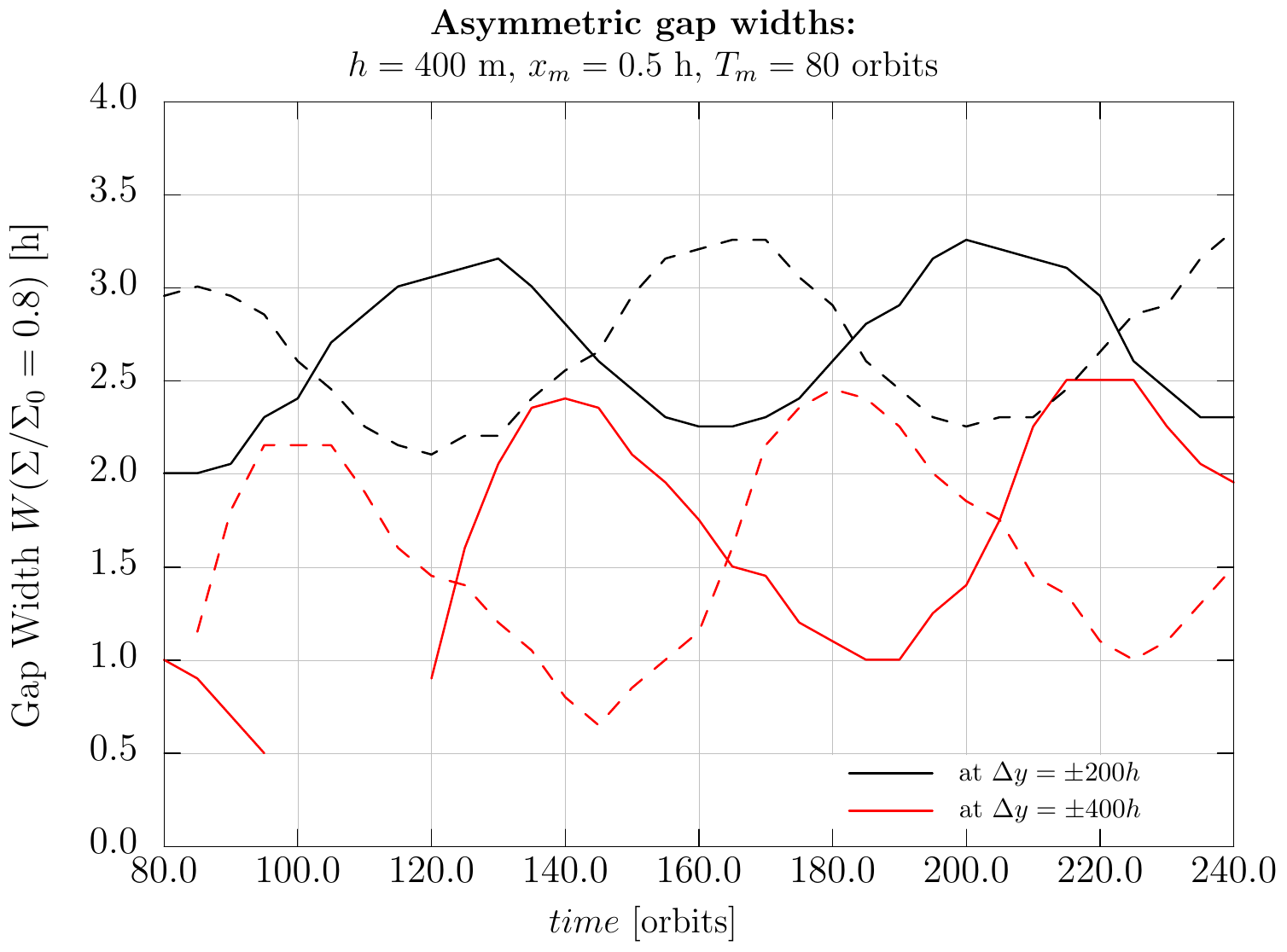}
\caption{\footnotesize Comparison of the gap width evolution of the 
inner (solid) and outer (dashed) propeller gap for a non-librating (left) 
and librating (right) central moonlet at $\Sigma/\Sigma_0=0.8$ for $\Delta y
\pm=\SI{200}{\hill}$ (black) and $\Delta y \pm \SI{400}{\hill}$ (red). 
For the non-librating moonlet the values of $W_s(t)$ fall on top of each 
other, giving mean values of $<W_s> \approx \SI{2.7}{\hill}$ for 
$\Delta y = \pm \SI{200}{\hill}$ and $<W_s> \approx \SI{1.9}{\hill}$ 
for $\Delta y = \pm \SI{400}{\hill}$, respectively. For the librating moonlet,
the gap widths oscillate with the moonlet and show an anti-phase relation. Their
mean values slightly differ: $<W_a(\Delta y = \pm\SI{200}{\hill})> =
\SI{2.76}{\hill}$ against $<W_a(\Delta y = \pm\SI{400}{\hill})> =
\SI{1.8}{\hill}$ for the inner gap and $<W_a(\Delta y = \pm\SI{200}{\hill})> =
\SI{2.63}{\hill}$ against $<W_a(\Delta y = \pm\SI{400}{\hill})> =
\SI{1.56}{\hill}$ for the outer gap. Interestingly, the amplitude of the
oscillation is increasing for larger azimuth.}
\label{fig:GapWidthEvo}
\end{figure*}

Figure \ref{fig:GapWidthEvo} shows a comparison of the time-evolution of the
asymmetric $W_{a}$ and symmetric $W_{s}$ gap widths at two fixed azimuthal 
distances ($\Delta y=\pm \SI{200}{\hill}$ and $\Delta y=\pm \SI{400}{\hill}$) 
and for the density level $\Sigma/\Sigma_0=0.8$.
The left panel shows the evolution of $W_s(t)$ for the symmetric propeller,
while the right one shows $W_a(t)$ for the asymmetric propeller.
For the symmetric case, the values of $W_s$ for the inner and outer gap fall 
on to of each other, while they increase in time to reach the saturation 
steady state which are $<W_s> \approx \SI{2.7}{\hill}$ for $\Delta y = \pm
\SI{200}{\hill}$ and $<W_s> \approx \SI{1.9}{\hill}$ for $\Delta y = \pm
\SI{400}{\hill}$. The latter lower value is to be expected because the 
viscous diffusion reduces the width $W_{s/a}(\Delta y)$ with growing 
azimuth $\Delta y$.

Quite different is the case for the librating moonlet, given in the right panel
of Figure \ref{fig:GapWidthEvo}, where the widths $W_a(t)$ oscillate with the
libration of the moonlet and are obviously in opposite phases (anti-phase). 
Again, the mean values $<W_a>$ are smaller for the larger azimuthal cut at 
$\Delta y = \pm \SI{400}{\hill}$, while, interestingly, the amplitude of the
oscillation is larger (about $<W_a(\Delta y = \pm\SI{200}{\hill})> =
\SI{2.76}{\hill}$ against $<W_a(\Delta y = \pm\SI{400}{\hill})> =
\SI{1.8}{\hill}$ for the inner gap and $<W_a(\Delta y = \pm\SI{200}{\hill})> =
\SI{2.63}{\hill}$ against $<W_a(\Delta y = \pm\SI{400}{\hill})> =
\SI{1.56}{\hill}$ for the outer gap). 

\subsubsection{Gap Depth} \label{sec:GapDepth}
Another measurable quantity is the gap depth, given by $d(t) =
\frac{\Sigma_{min}(r) - \Sigma_0}{\Sigma_0}$.
As in the asymmetric case the gap widths $W(t)$ and -lenghts $L(t)$ 
are changing with time, an analogous behavior is to be expected for $d(t)$.

\begin{figure*}
\centering
\includegraphics[width=0.49\textwidth]{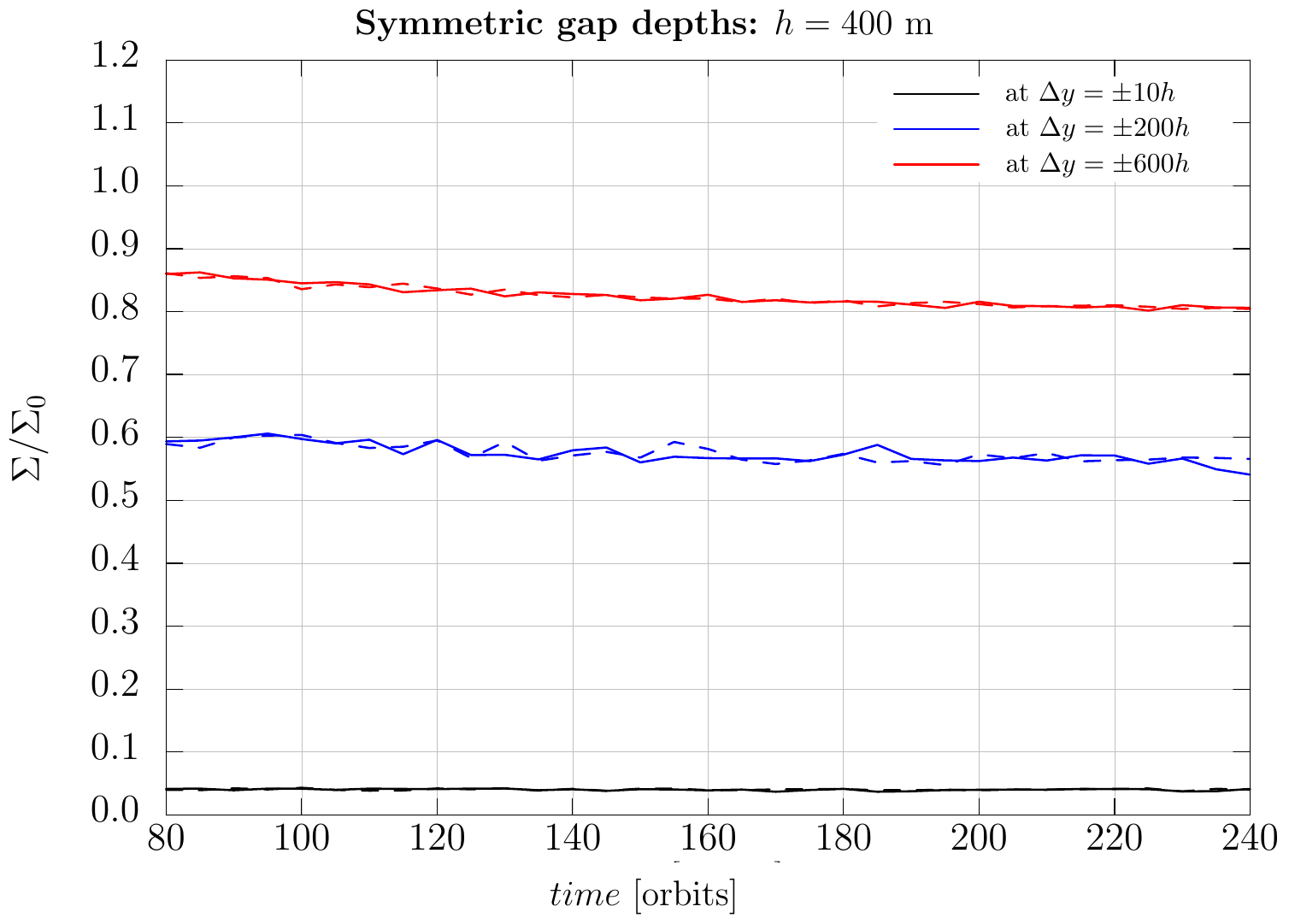}
\includegraphics[width=0.49\textwidth]{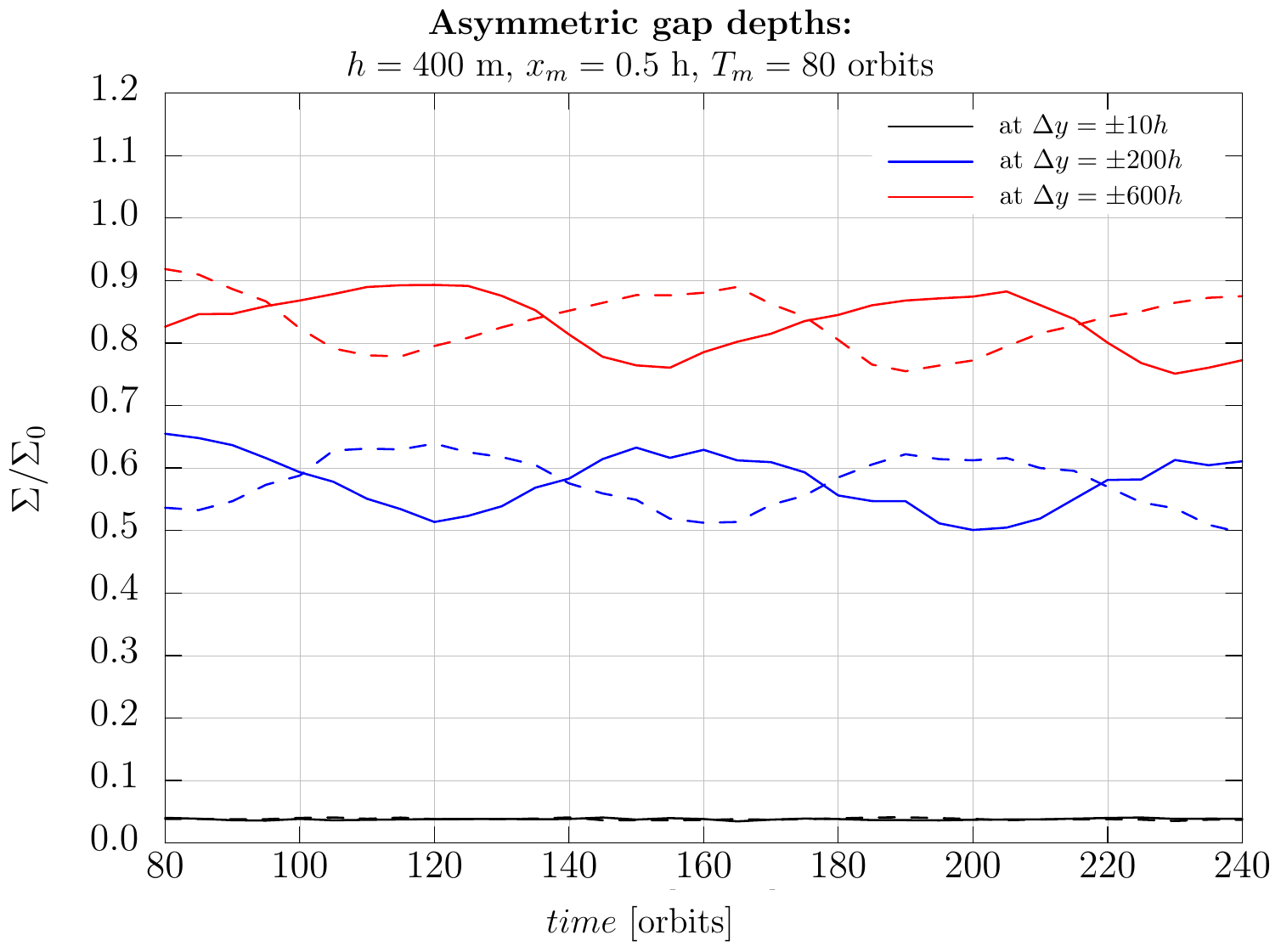}
\caption{\footnotesize Evolution of the propeller gap depth for the symmetric
(left) and asymmetric (right) propeller at different azimuthal positions to
the moonlet $\Delta y=\pm\SI{10}{\hill}$ (black), 
$\Delta y = \pm\SI{200}{\hill}$ (blue) and $\Delta y=\pm\SI{600}{\hill}$ 
(red). The solid and dashed lines represent the inner and outer propeller gap.}
\label{fig:GapDepthEvo}
\end{figure*}

Figure \ref{fig:GapDepthEvo} shows the time-evolution of $d(t)$ for the
symmetric case (left panel) and the asymmetric case (right panel) for different
values of $\Delta y$.

While for the symmetric case (left panel) the values of $d_s(t)$ for the outer 
and inner propeller wing coincide and slightly relax to the steady state, the 
asymmetric value of $d_a(t)$ (right panel) oscillates with the moonlet libration,
similarly to the width $W_a(t)$. The amplitude of this depth-variation is about
$\Delta d(t) = \frac{d_{max}-d{min}}{<d>} \approx 10\%$.

\subsection{Azimuthal Gap Relaxation}
To estimate the azimuthal gap relaxation, we study the dependence of the 
gap minimum on the azimuthal distance to the moonlet. 

\begin{figure*}
\centering
\includegraphics[width=0.49\textwidth]{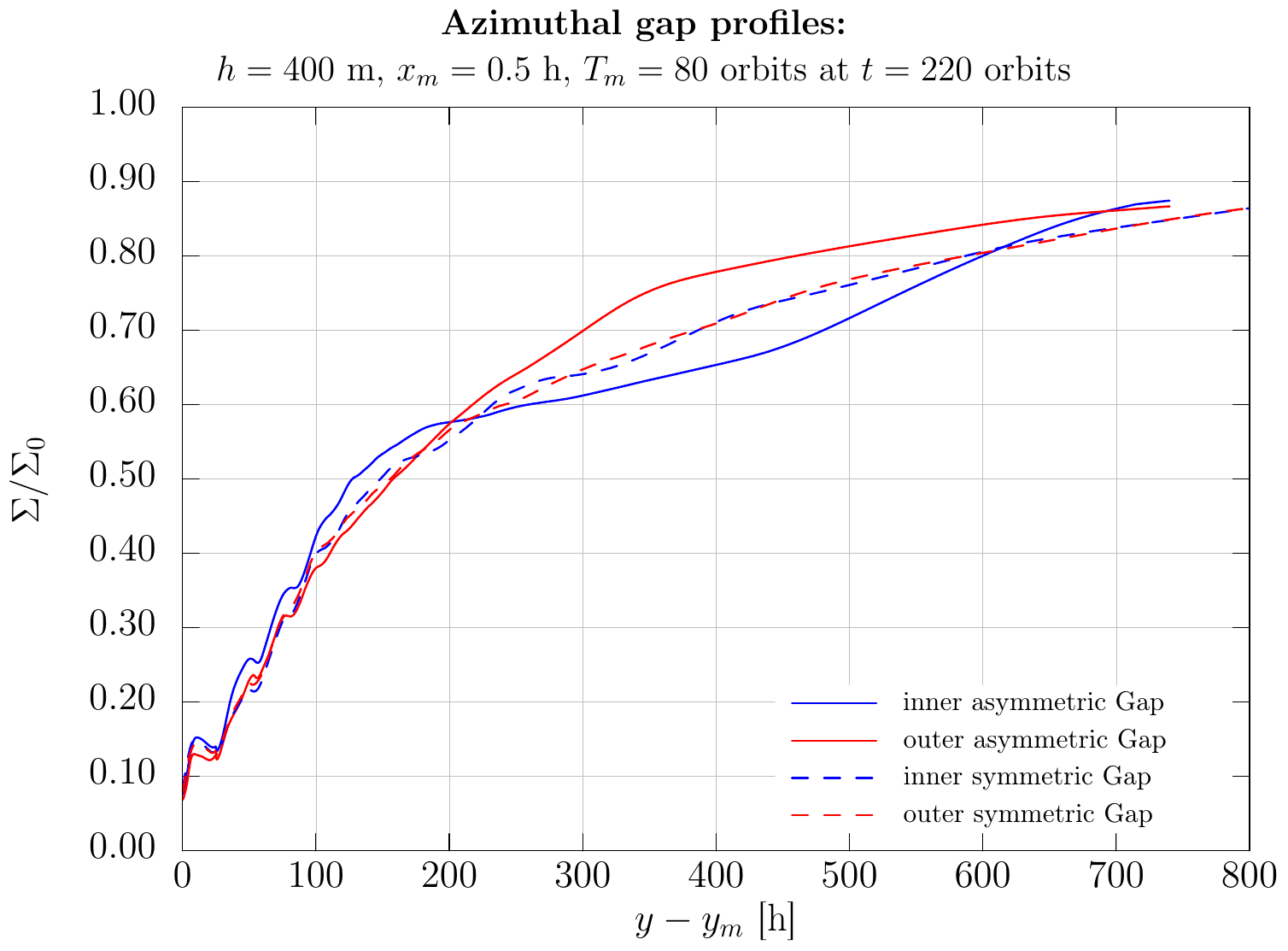}
\includegraphics[width=0.49\textwidth]{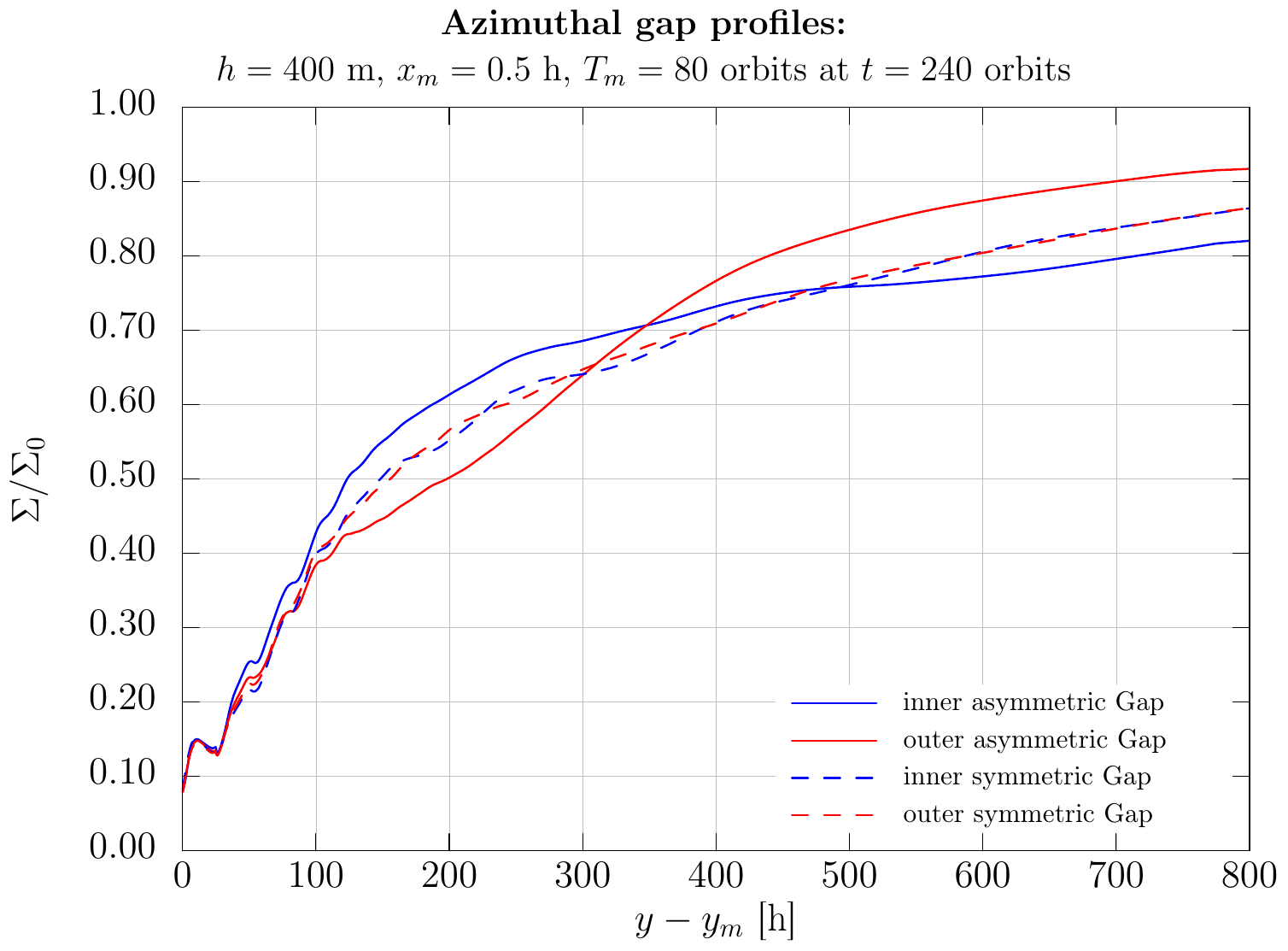}
\caption{\footnotesize Azimuthal gap relaxation profile for the symmetric
(dashed) and asymmetric (solid) gap structure at $T=\SI{220}{\orbits}$ (left) 
and $T=\SI{240}{\orbits}$. Comparing the inner (blue) and outer (red) gap 
profiles in the symmetric case, we clearly see how both profiles match 
perfectly. In the asymmetric case the symmetry of the propeller is broken, 
which can be seen from the different lengths and the different depths of the 
gaps at different azimuthal distances to the moonlet.}
\label{fig:gap_relaxation}
\end{figure*}

In order to extract the azimuthal gap profiles, we smoothen the radial 
profiles with a moving box averaging method, where the box size has 
been set to $\Delta x = \pm \SI{0.5}{\hill}$. This reduces the radial 
scattering of the found gap minima locations especially in the wake region. 
Further, we use another moving box averaging process to avoid the azimuthal 
scattering of data due to the wakes. For this reason, we set the box size to 
$\Delta y = \pm \SI{25}{\hill}$. 
The resulting azimuthal gap relaxation profiles are shown in Figure
\ref{fig:gap_relaxation} for the symmetric (dashed lines) and asymmetric 
(solid lines) propeller.

In the left and right panel the asymmetric propeller minima are presented at 
orbit 220 and 240 of the integration time, respectively. 
The red and blue colors denote the outer and inner gaps. Comparing the 
non-librating and librating moonlet a clear difference in the 
gap relaxation can be observed. While for the symmetric propeller the inner 
and outer gaps are closing in the same way with growing azimuth 
(except small variations due to the wakes and noise), the gap depths change
along the azimuth for the asymmetric propeller. 

Due to the retardation, the influence of changes in the moonlet's 
radial libration are found by the intersection points of the minima density 
curves (see right panel, Figure \ref{fig:gap_relaxation}) for the inner and 
outer gap. As an example, seen in the right panel, the inner gap is closing 
faster than the outer one until \SI{400}{\hill} azimuthal distance to the 
moonlet, respectively. There, an intersection point of both gap profile 
curves can be found. From this distance on, the outer gap is closing faster 
and stays over the density minima of the inner gap. 
This intersection point is caused at $T\approx\SI{200}{\orbits}$, 
where the moonlet is at its lowest radial elongation $x_m=\SI{-0.5}{\hill}$ 
(compare with Figure \ref{fig:RadialProfileEvo}). At this turning point, 
the moonlet starts to migrate outwards again.

\subsubsection{Gap Length} \label{sec:gap_length}
From the azimuthal gap profile we define the gap length $L_{80}$ as the
azimuthal distance to the moonlet at $\Sigma/\Sigma_0 = 0.8$ for the inner and
outer gap structure. 

\begin{figure*}
\centering
\includegraphics[width=0.49\textwidth]{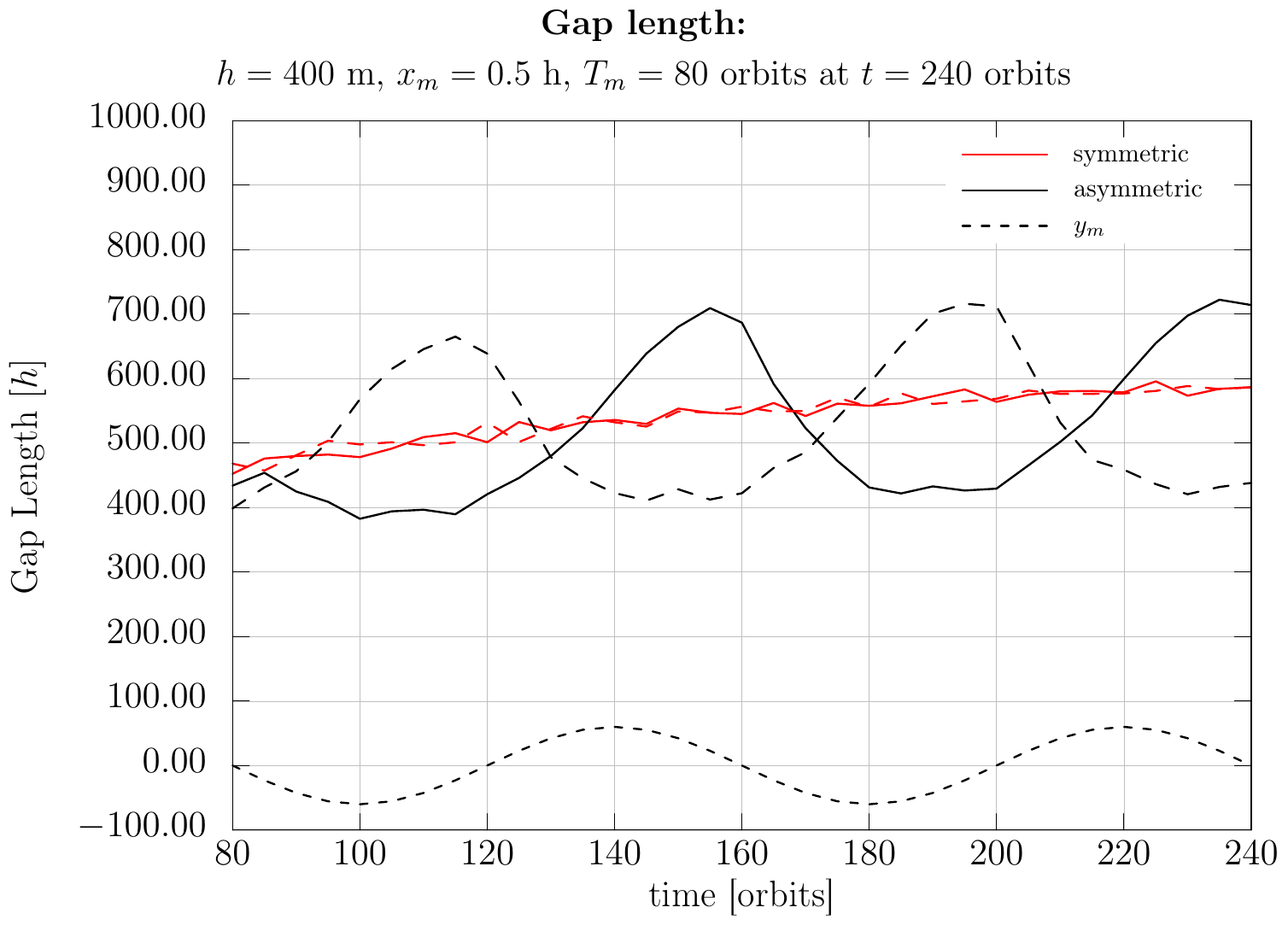}
\caption{\footnotesize Comparison of the gap length at 80\% 
relaxation for the symmetric (red) and asymmetric propeller (black). 
The length of the inner and outer gap are given by the solid and dashed lines.
The results have been obtained from the simulation of a $h=\SI{400}{\meter}$
sized moonlet. For the asymmetric propeller the moonlet was librating with a
radial amplitude of \SI{0.5}{\hill} and period of \SI{80}{\orbits}. For
comparison the azimuthal evolution of the librating moonlet is plotted as the
black dotted line as well. For both profile plots a moonlet-centered reference
frame has been chosen for better comparison.}
\label{fig:GapLengthEvo}
\end{figure*}

The evolution of the gap length $L_{80}(t) = L(\Sigma/\Sigma_0 = 0.8,t)$ is 
presented in Figure \ref{fig:GapLengthEvo}, where the symmetric (red color) 
and asymmetric (black color) propeller is presented. The dashed and solid 
lines denote the outer and inner gaps. The dotted line represents the 
azimuthal evolution of the azimuthal moonlet position as a reference. 
The asymmetric propeller gap length varies in phase with the moonlet motion. 
For the non-librating central moonlet, the mean gap length $<L_{80}>$ is 
about \SI{539.5}{\hill}, respectively, while the for asymmetric propeller, 
the mean gap lengths slightly differ ($<L_{80}> = \SI{511}{\hill}$ for the inner
gap against $<L_{80}>=\SI{523}{\hill}$ for the outer gap) and the maximum 
difference of the gap lengths is about \SI{300}{\hill}.

Although the depths of the gaps only differ by about 10\% (compare Section
\ref{sec:GapDepth}) these variations result in large azimuthal changes in the 
gap lengths. Thus, studying the azimuthal gap profiles is the favorable
method to search for the imprint of the asymmetry.

\subsection{The Gap Contrast}
As already carried out, the azimuthal gap profiles of the inner and outer gap
structures differ for a librating moonlet. In order to extract the 
perturbation by the moonlet motion and to study its propagation through the
gaps, we define the \emph{gap contrast} as the difference of the inner and 
outer gap profiles. 

The retardation of the symmetry-breaking perturbation by the moonlet libration
along the gap length sets a memory timescale $T_{gap}$. Changes in the radial
moonlet motion, happening within this timescale, become visible as zero value
crossings and help to differ a periodic motion from a migration in this way.

\begin{figure*}
\centering
\includegraphics[width=0.49\textwidth]{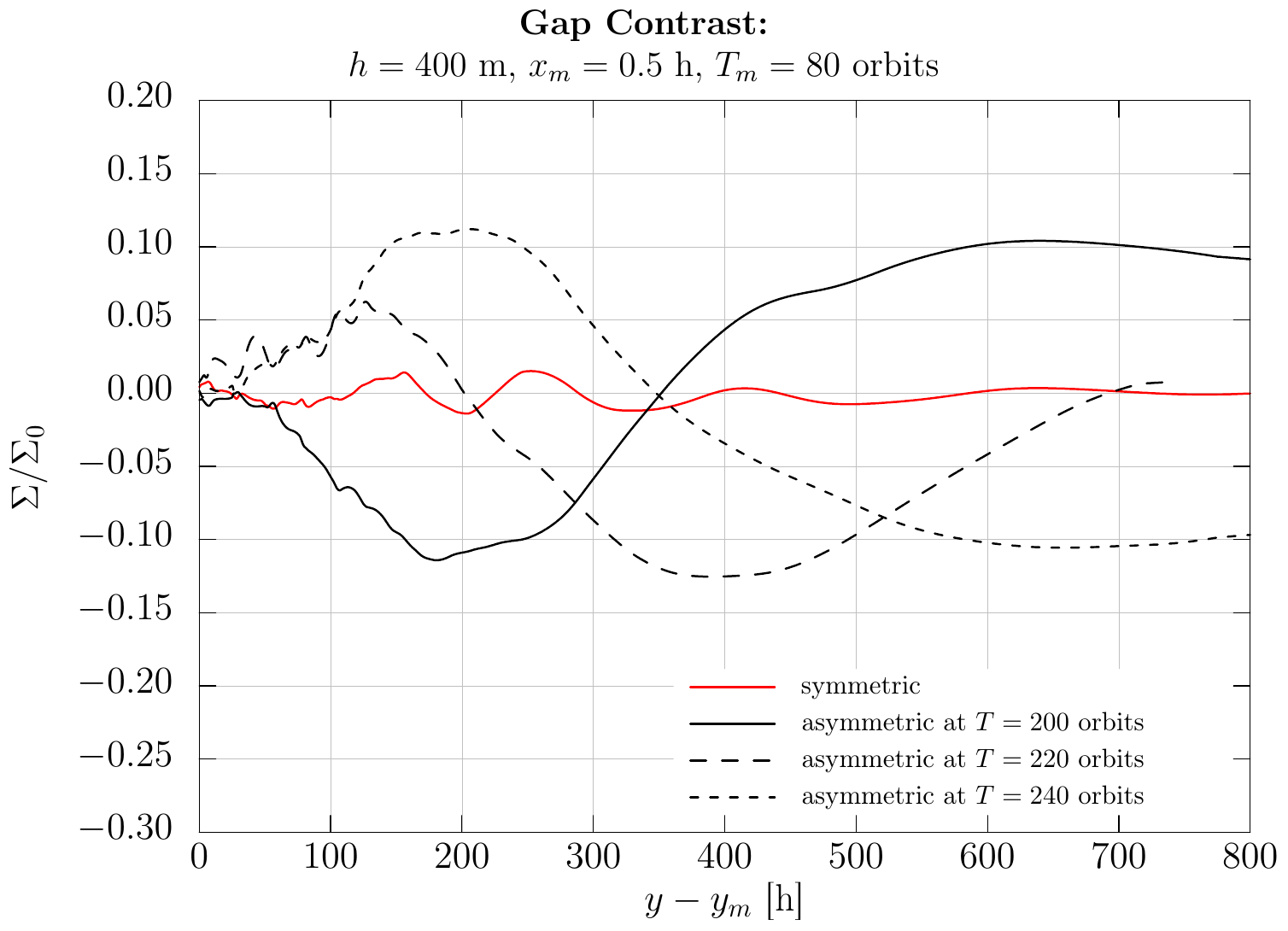}
\caption{\footnotesize Difference of the azimuthal gap relaxation of
the inner and outer gap (\emph{gap contrast}) for a non-librating (red) and 
librating (black) central moonlet of size $r_h = \SI{400}{\meter}$ at orbits
\num{200}, \num{220} and \num{240} (solid, dashed and dotted).}
\label{fig:contrast}
\end{figure*}

Figure \ref{fig:contrast} shows the gap contrast estimated from the azimuthal 
gap relaxation profiles presented in Figure \ref{fig:gap_relaxation}.
The red solid line represents the gap contrast of the symmetric propeller
structure along increasing azimuthal distance to the moonlet. 
Its maximum corresponds to deviations caused by the wake region 
($\Sigma/\Sigma_0 \approx 2\times 10^{-2}$) and thus defines the level, 
where the symmetric and asymmetric propeller will not be distinguishable anymore.
The black lines in Figure \ref{fig:contrast} represent the gap contrast for the
asymmetric gap structure at different integration times, where the solid, 
dashed and dotted lines show the gap contrast at $T=$\num{200}, \num{220} and
\SI{240}{\orbits}. The different snapshots for the asymmetric 
propeller demonstrate, how the perturbation by the moonlet libration propagates 
through the gaps. As an example, the minimum at $T=\SI{200}{\orbits}$ shifts 
from $\Delta y = \SI{200}{\hill}$ and $\Sigma/\Sigma_0=\num{-0.1}$ to 
$\Delta y = \SI{400}{\hill}$ and $\Sigma/\Sigma_0=\num{-0.12}$ at orbit
\num{220} and further to $\Delta y = \SI{650}{\hill}$ and 
$\Sigma/\Sigma_0=\num{-0.1}$ at orbit \num{240}. The transport of 
perturbations along the azimuth resembles a propagating wave package. Like a
dissipating wave package, the perturbation gets smeared along the azimuth due 
to diffusive effects.

\section{Ring - Moonlet Interactions} \label{sec:ring_gravity}
The symmetry breaking of the propeller structure disturbs the force balance of
the gravity by the ring ensemble reacting on the moonlet 
\citet{Seiler2017ApJL}. Here, we calculate the resulting force of the 
surrounding ring material on the moonlet. Therefore, we calculate the 
gravity of every grid cell onto the moonlet and sum over all the contributing 
cells of the grid. We exclude a circular area 
$r=\left[\left(x-x_m\right)^2 + \left(y-y_m\right)^2\right]^{1/2}=\SI{1.2}{\hill}$ 
around the moonlet from the force calculation in order to account for the 
physical dimension of the moonlet, which is treated as a point mass in our 
simulations. The resulting total gravitional interaction between the ring and
the moonlet $F_x$ and $F_y$ in radial and azimuthal direction is given by the sum 
of the contributing cells 

\begin{align}
  F_x \left(x,y,x_m,y_m \right) & = -G \Delta A
  \sum_{\text{Grid}} \Delta \Sigma_i \frac{x_m-x_i}{\left|\vec{r}_m-\vec{r}_i\right|^3} \\
  F_y \left(x,y,x_m,y_m \right) & = -G \Delta A
  \sum_{\text{Grid}} \Delta \Sigma_i \frac{y_m-y_i}{\left|\vec{r}_m-\vec{r}_i\right|^3} \, \text{,}
\end{align}

with $\vec{r_i} = (x_i,y_i)$ and $\vec{r}_m = (x_m,y_m)$ the position vectors 
of the i-th cell and the moonlet. The quantities 
$\Delta \Sigma_i = \Sigma_i - \Sigma_0$ and 
$\Delta A = \Delta x_i \times \Delta y_i$ denote the difference in the surface
mass density with respect to the intial one and the surface of the 
equal-sized grid cells.

\begin{figure*}
\centering 
\includegraphics[width=0.49\textwidth]{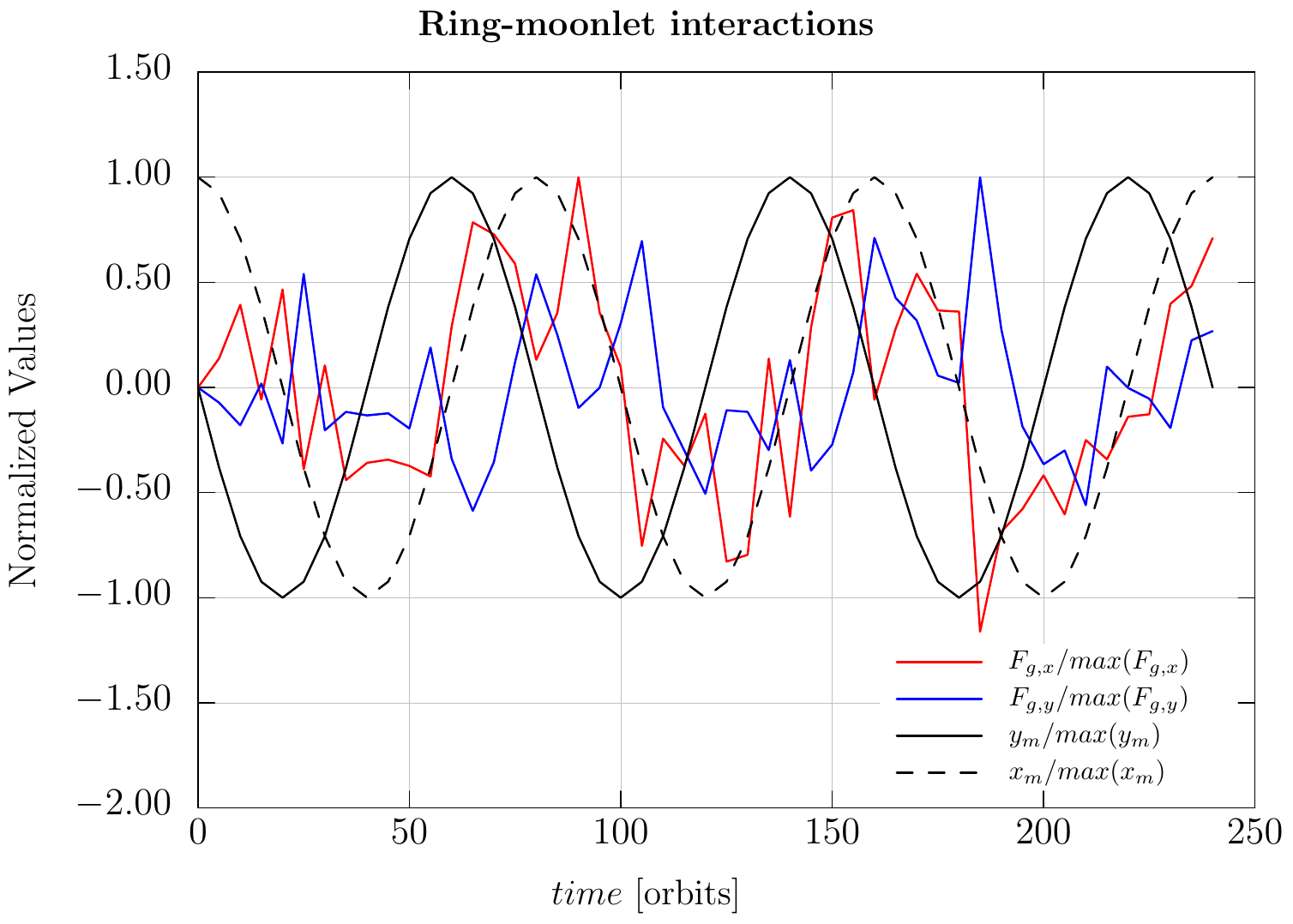}
\includegraphics[width=0.49\textwidth]{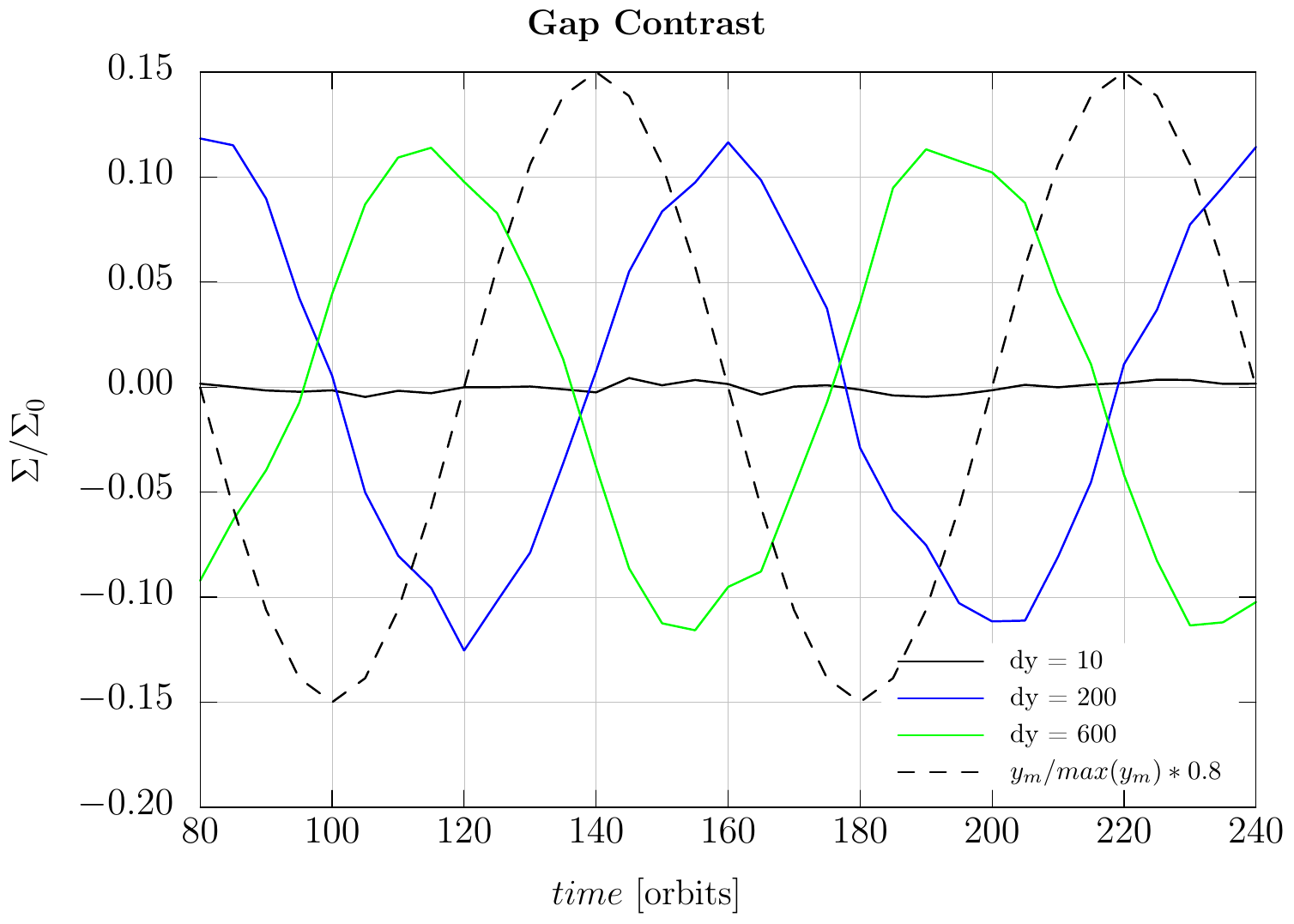}
\caption{\footnotesize Left: 
Time evolution of the normalized moonlet motion in x (black dashed line) and y
(black solid line) direction and the x (blue) and y (red) component of the ring
gravity acting on the moonlet. 
Right: Evolution of the gap contrast at different azimuthal positions,
illustrating, how the variation of the gap contrast is following the moonlet
motion (black dashed line).}
\label{fig:forces}
\end{figure*}

The left panel in Figure \ref{fig:forces} illustrates the evolution of the 
normalized ring gravity, where the x (red) and y (blue) component of the ring 
gravity are following the moonlet motion ($x_m$ and $y_m$ are represented by 
the dashed and solid black lines). The right panel shows a comparison of the gap
contrast evolution and the moonlet evolution, where the gap contrast is
presented at three different azimuthal distances to the moonlet. 
The evolution for $\Delta y = \pm\SI{200}{\hill}$ and 
$\delta y=\pm\SI{600}{\hill}$ displays the retardation set by the Kepler shear 
and the decreasing amplitude of the gap contrast with increasing azimuths. 
The force and the maximum contrast are following the moonlet motion with a 
slight delay (\emph{phase shift}).

\section{Systematic Asymmetry Predictions} \label{sec:predictions}
In this section we test the dependence of the propeller's asymmetry on the 
\emph{a)} the libration amplitude and \emph{b)} the libration period. This 
allows us to make predictions for the observability of the asymmetry of 
propeller gap structures in Saturn's rings. Therefore, at first, we keep 
the libration period constant at $T_{m}=\SI{80}{\orbits}$ and vary the 
initial radial amplitude $x_{m,0}$. In a second analysis, we perform 
simulations, where the initial radial amplitude $x_{m,0}=\SI{1}{\hill}$
will be kept constant, while the libration period is varied. For both 
scenarios we will study the gap contrast, the gap length and width at 
$\Sigma/\Sigma_0=0.8$.

Note, that our analysis does not consider any resonances in the vicinity of 
the propeller structure, which might have an additional effect of the final 
asymmetry for larger radial amplitudes and libration periods. 

For each simulation parameter, at each time the above quantities will be 
determined. Further, the mean value, the maximum and minimum at each time will 
be estimated along the azimuth. Finally, for the complete simulation time 
frame the global maximum, minimum and mean values will be calculated. 
The resulting values will be presented in the following sections.

\subsection{Dependence of Asymmetry on the Libration Amplitude}
\label{sec:asym_vs_ampl}
We vary the libration amplitude $x_{m,0}$ from \SI{0.125}{\hill} through 
\SI{4}{\hill}. 

\begin{figure*}
\centering
\includegraphics[width=0.49\textwidth]{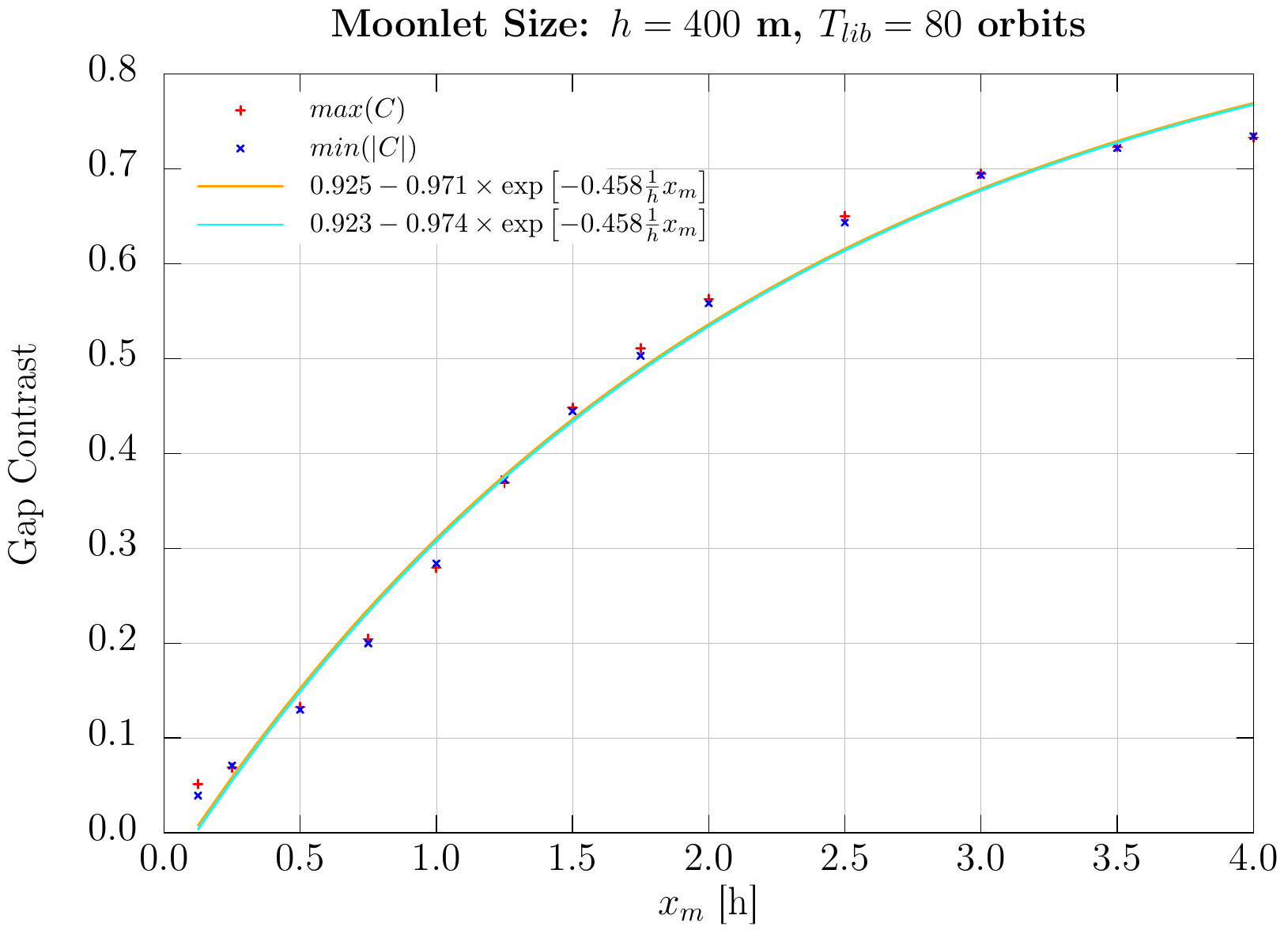}
\caption{\footnotesize Dependence of the maximal asymmetry from the moonlet's 
libration amplitude for constant libration period $T_m=\SI{80}{\orbits}$.}
\label{fig:ContrastVsAmpl}
\end{figure*}

In Figure \ref{fig:ContrastVsAmpl} the dependence of the maximum gap contrast 
from the radial amplitude of the moonlet libration is shown. Here, we can 
fit the data by the exponential relation 

\begin{equation}
  C \left(x_m\right) = C_0 - C_{max} \exp \left[ -\lambda x_m \right] \, ,
\end{equation}

with $C_{i,0} = 0.925$, $C_{i,max} = 0.971$ and 
$\lambda_i = \SI{0.458}{\per\hill}$ for the maximum 
$C_{o,0} = 0.923$, $C_{o,max} = 0.974$ and 
$\lambda_o = \SI{0.458}{\per\hill}$ for the minimum gap contrast.

Starting from small amplitudes the contrast is increasing for growing initial 
amplitude until the saturation of about $C=0.925$ is reached. From this 
amplitude on the gap contrast is not increasing anymore. For amplitudes larger 
$x_{m,0}=\SI{1.6}{\hill}$ the moonlet starts to migrate through its created 
gaps, and thus, destroys its own propeller structure, explaining the 
maximum saturation level. 

\begin{figure*}
\centering
\includegraphics[width=0.49\textwidth]{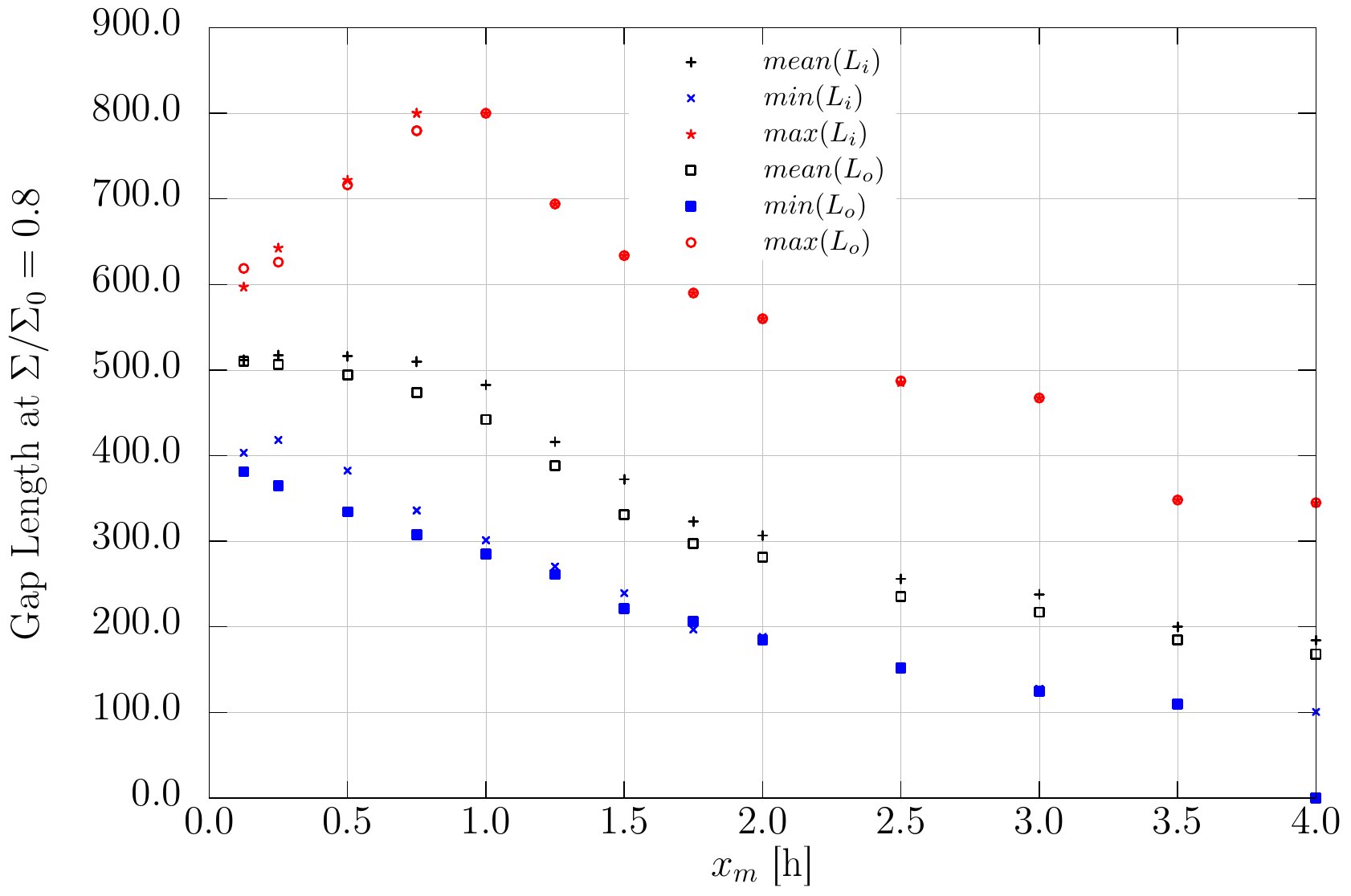}
\includegraphics[width=0.49\textwidth]{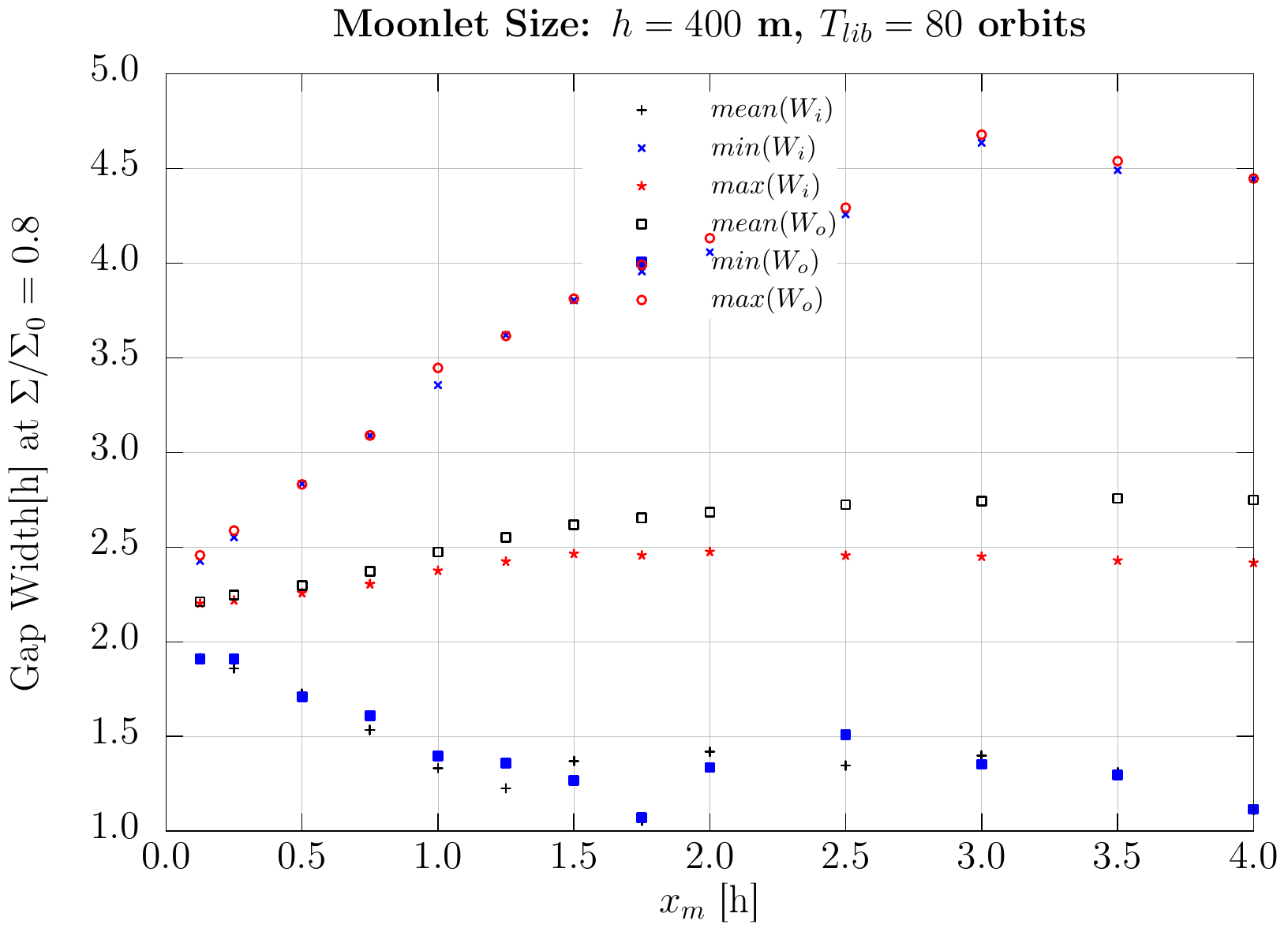}
\caption{\footnotesize Dependency of the gap lenght (left) and gap width (right) 
at $\Sigma/\Sigma_0 = 0.8$ from the moonlet's libration amplitude for the 
inner and outer gap.}
\label{fig:WidthVsAmpl}
\end{figure*}

Figure \ref{fig:WidthVsAmpl} illustrates the dependence of the gap length
$L_{80}$ (left panel) and the gap width $W_a$ (right panel) on the libration 
amplitude. For the gap length, a clear drop in the mean gap length can be noticed 
for larger amplitudes, while the maximum difference in the gap lengths of the 
inner and outer gap first increases from \SI{200}{\hill} to \SI{500}{\hill} 
for $x_{m,0}=\SI{0.125}{\hill}$ to $x_{m,0}=\SI{1}{\hill}$ and then reduces to 
about \SI{200}{\hill}. With the gap length the gap width changes as well 
(see right panel of Figure \ref{fig:WidthVsAmpl}), but here, the mean gap width 
is increasing from \SI{2.25}{\hill} to \SI{2.6}{\hill} for larger amplitudes, 
until at about $x_{m,0}=\SI{1.6}{\hill}$ a saturation level is reached. 
A similar result can be seen for the maximum difference of the inner and outer 
gap width, which increases from \SI{0.5}{\hill} to \SI{3.5}{\hill}, respectively.

To summarize, with larger radial amplitudes the perturbation by the moonlet
motion increases, which results in shorter gap lengths, broader gaps and larger 
gap contrast until the perturbation exceeds a critical level, where the
creation of the propeller structure is prevented.

\subsection{Dependence of Asymmetry on the Libration Period}
\label{sec:asym_vs_T}
We vary the libration period of the moonlet from $T_{m}=\SI{15}{\orbits}$ 
to $T_{m}=\SI{320}{\orbits}$, while we keep the radial amplitude fixed at 
$x_{m,0}=\SI{1}{\hill}$.

\begin{figure*}
\centering
\includegraphics[width=0.49\textwidth]{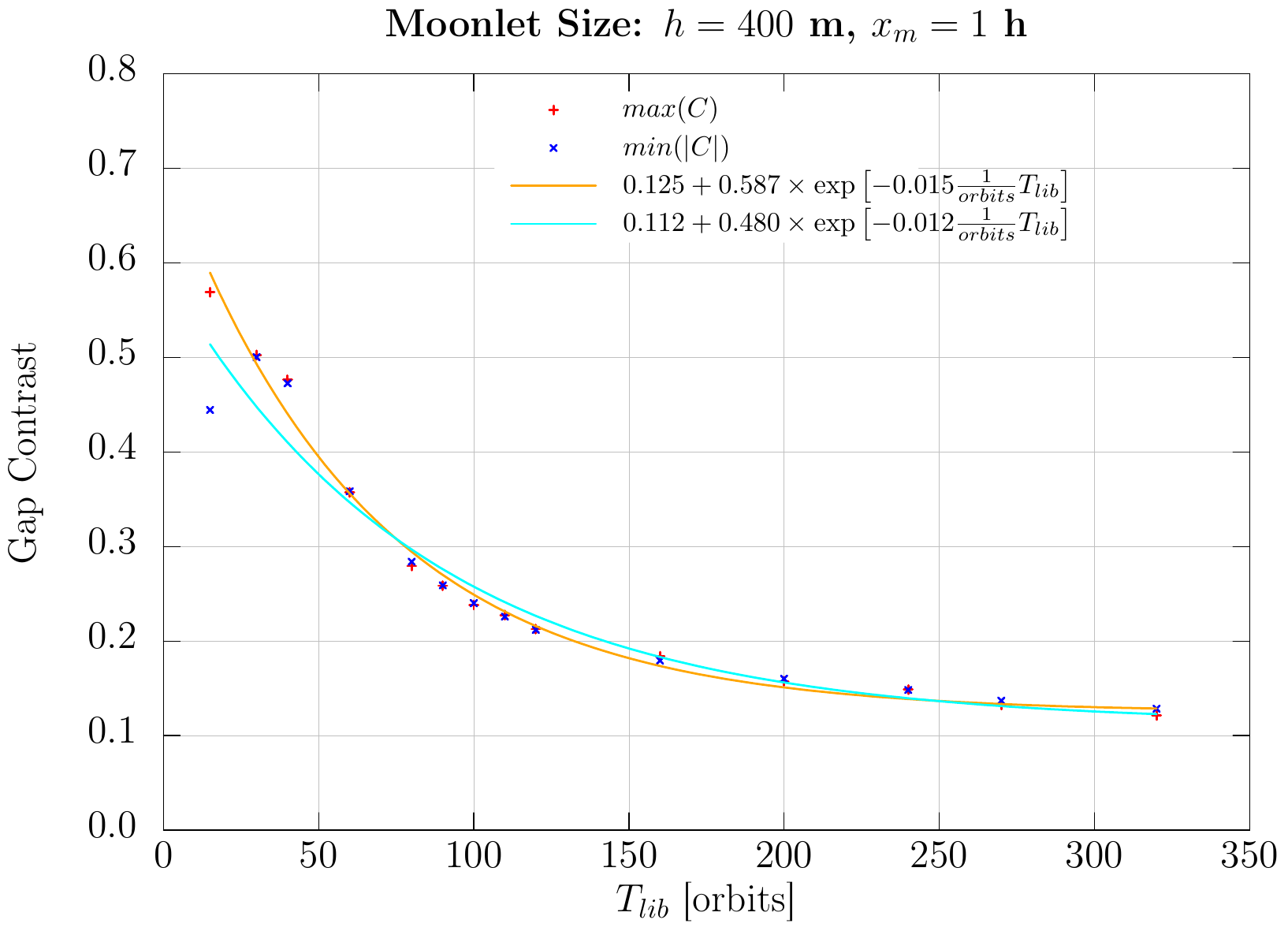}
\caption{\footnotesize Dependence of the maximal asymmetry on 
the moonlet's libration period for constant radial amplitude 
$x_{m,0}=\SI{1}{\hill}$.}
\label{fig:AsymVsT}
\end{figure*}

Figure \ref{fig:AsymVsT} shows the dependence of the gap contrast on the 
libration period, where an exponential relation given by 

\begin{equation}
  C \left(T_{lib}\right) = C_0 + C_{max} \exp \left[ -\lambda
  T_{lib} \right] \, ,
\end{equation}

with $C_0 = \num{0.125}, \, C_{max} = \num{0.587}$ and 
$\lambda = \SI{0.013}{\per\orbit}$ can be found.

For fixed radial amplitude the changes for larger libration periods result 
in a growing azimuthal excursion of the moonlet (see Eq. 
\ref{equ:moonlet_motion}). As the libration period grows, the relative 
velocity to the unperturbed orbit decreases and thus the perturbation by 
the changing moonlet position can be easily transported to larger azimuthal 
distances, resulting in a more symmetric propeller structure and therefore in a
lower gap contrast (see Figure \ref{fig:AsymVsT}). 

\begin{figure*}
\centering
\includegraphics[width=0.49\textwidth]{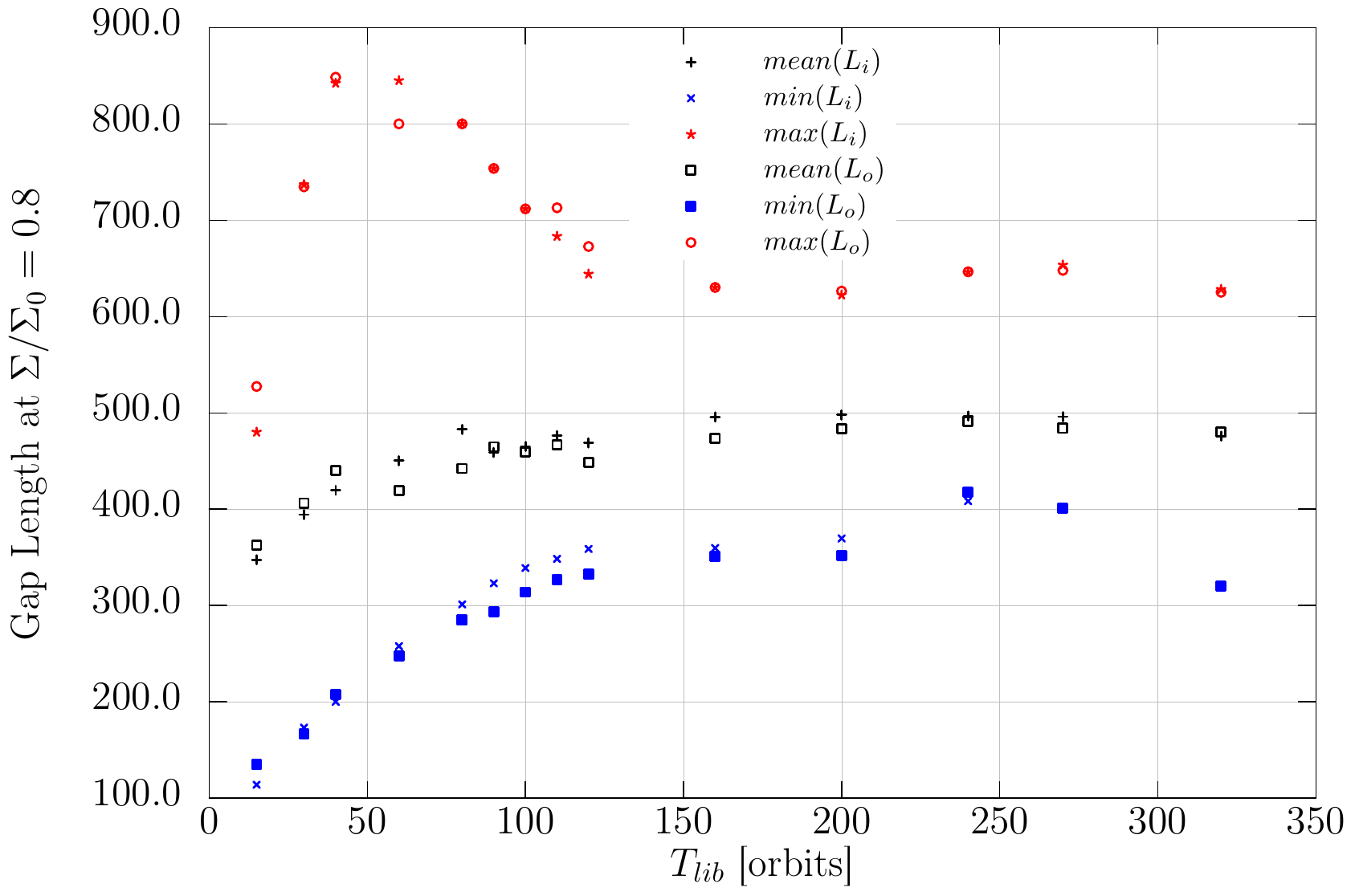}
\includegraphics[width=0.49\textwidth]{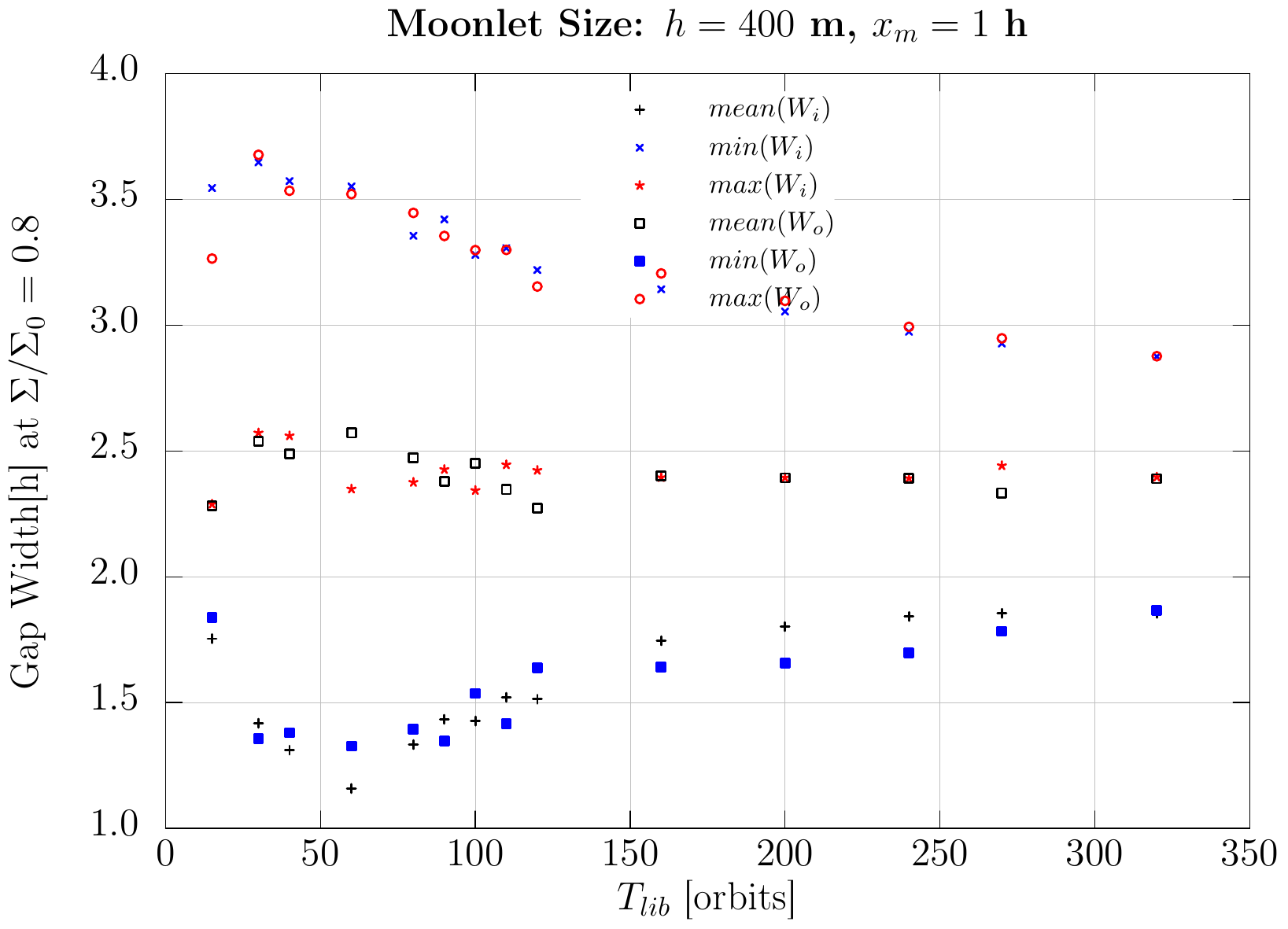}
\caption{\footnotesize Dependence of the gap length (left) and gap width (right) 
at 80\% gap relaxation from the moonlet's libration amplitude for fixed radial
amplitude $x_{m,0}=\SI{1}{\hill}$. }
\label{fig:WidthVsT}
\end{figure*}

In Figure \ref{fig:WidthVsT} the dependence of the gap length (left panel) and
the gap width (right panel) on the libration period is presented.

The gap length (left panel) first increases from \SI{350}{\hill} to about 
\SI{500}{\hill} with respect to the libration period interval from 
$T_0=\SI{15}{\orbits}$ through $T_0=\SI{80}{\orbits}$. For this interval, 
the maximum difference in the gap lengths between outer and inner gap 
increase from \SI{400}{\hill} to about \SI{600}{\hill}. 
At $T_m=\SI{80}{\hill}$ the gap passage time equals the libration period of 
the moonlet allowing the perfect transport of the perturbation through the gaps.
At this optimal parameter set the gap width reaches its maximum of about 
\SI{2.5}{\hill}, respectively, while the differences in the gap width between 
inner and outer gap are at their maximum of about \SI{2}{\hill} 
(compare right panel). For increasing libration periods, the relative velocity 
of the moonlet to its unperturbed orbit decreases, resulting in smaller 
perturbations and therefore in a more symmetric propeller structure. For this 
reason, the mean gap length drops to about \SI{500}{\hill} -- close to the value 
of the unperturbed propeller gap length. Further, the maximum difference in 
the gap length drops to an almost constant level of \SI{200}{\hill} as well.
Exceeding a libration period of $T_m=\SI{80}{\orbits}$ the mean gap width stays
almost constant at \SI{2.5}{\hill}, which is very close to the unperturbed value
of \SI{2.6}{\hill}. Nevertheless, the moonlet motion still induces a difference
of the gap widths of about \SI{1}{\hill}.

To summarize, for small libration periods ($T_m\le T_{gap}$) the asymmetry is 
large because of the strong perturbation by the moonlet motion and the
retardation, while for large periods $(T_m > T_{gap}$) the asymmetry gets 
smaller due to quasi-static changes in the moonlet motion. However, the
asymmetry is still measurable but depends on the actual libration phase angle of
the moonlet at the observation time.

\subsection{Application to the Propeller \bleriot} 
\label{sec:application_bleriot}
The excess motion of \bleriot{} has been reconstructed by three overlaying
harmonic functions, where we assume the moonlet to librate with the largest 
amplitude and period of about \SI{1845}{\kilo\meter} and \num{11.1} years 
\citep[see e.g.][]{Seiler2017ApJL,Spahn2018Book}. Translating the fitted 
period to number of orbits results 
$T_m = \SI{11.1}{\years} = \SI{6941}{\orbits}$ and thus, the radial amplitude
for the moonlet is given by 

\begin{equation}
  \left| \delta x \right| = \frac{2}{3} \frac{\Delta y}{\Omega t} = 
    \frac{2}{3} \frac{\SI{1845}{\kilo\meter}}{\SI{6941}{\orbits}} =
    \SI{0.177}{\kilo\meter} 
\end{equation}

This agrees with the radial amplitude estimated from orbital fits by
\citet{Spahn2018Book}, who estimated the radial deviation to be about
\SI{200}{\meter}. The Hill radius of \bleriot{} is about 
$r_H=\SI{450}{\meter}$ \citep{Hoffmann2016DPS,Seiler2017ApJL} and thus, the 
estimated radial amplitude corresponds to $x_{m,0}\approx\SI{0.4}{\hill}$.

For \bleriot{} \citet{Seiler2017ApJL} estimated $T_{gap} \approx \num{0.5}$ 
years which corresponds to \SI{313}{\orbits}, respectively. Thus, for 
\bleriot{} we need to consider the case 
$T_m = \SI{11.1}{\years} = \SI{6941}{\orbits} \gg \SI{313}{\orbits} =
\SI{0.5}{\years} = T_{gap}$. At this ratio the gap contrast is already at its 
lowest level of about $0.125$ in comparison to the noise level of the 
unperturbed propeller gap contrast of $0.02$. 

The maximum gap contrast illustrated Figure \ref{fig:ContrastVsAmpl} can be
linearly approximated in the range $x_{m,0} = [0,\SI{1.5}{\hill}]$. 
Therefore, the expected gap contrast level of about $C_0=0.125$ estimated 
for $x_{m,0}=\SI{1}{\hill}$ (see Figure \ref{fig:AsymVsT}) can be 
interpolated to $x_{m,0} = \SI{0.5}{\hill}$, yielding \num{0.063}, respectively.
Following this interpolation, the maximum variation in gap length and gap 
width are expected to be about \SI{125}{\hill} and \SI{0.5}{\hill},
respectively.

\section{Conclusion and Discussion} \label{sec:discussion}
In this work, we have studied the formation of asymmetric propeller structures 
in Saturn's A ring, assuming a libration of the moonlet.
Note, that in this work, we did not consider the reason for the moonlet 
libration, which we simply considered as given. In pratice, librational behavior
sensitively depends on the perturbation forces and sources (e.g. resonances,
density fluctuations, particle collisions).
It turned out, that the additional moonlet motion is perturbing the induced
propeller structure, causing an asymmetry in this way. 
The perturbation by the moonlet motion gets transported through the gaps 
with the Kepler speed and thus arrives at the gap ends after a delay time 
$T_{gap}$. This retardation depends on the gap length and sets a memory time, 
which finally causes the asymmetric appearance of the propeller structure in 
the images. 
The main implications of our simulations are: 

\subsection{Timescales vs Asymmetry}
We studied the dependence of the asymmetry on the libration period and 
amplitude. Three general cases have been investigated:

\begin{itemize}
\item[i)] $T_m < T_{gap}:$ For small libration periods the asymmetry is large.
This results from the strong perturbation by the moonlet motion, which causes
large variations in all gap properties, which even can prevent the propeller 
formation. Thus, small libration periods are similar to the scenario of large 
libration amplitudes.
\item[ii)] $T_m = T_{gap}:$ The strongest asymmetry has been identified, if 
the libration period matches the gap timescale. In this case, the perturbation
by the moonlet motion gets transported in the most effective way (
in a kind synchronized way), resulting in the largest variations in the gap 
length and width.
\item[iii)] $T_m > T_{gap}:$ The quasi-static changes in the moonlet motion
result in a less asymmetric shape of the propeller. Thus, the gap contrast 
is minimal and the variations in the gap width and gap length are rather small.
\end{itemize}

\subsection{Libration vs. Migration}
The retardation limits the observability of the asymmetry and with it the
chance to distinguish a migration of the moonlet from a libration. 
Changes in the direction of the moonlet motion (inward and outward movement) 
result in zero value crossings in the gap contrast along the azimuth. Depending
on the libration period, three limiting cases can be distinguished:

\begin{itemize}
\item[i)] $T_m < T_{gap}:$ The frequent changes in the moonlet motion can be
seen due to the retardation, resulting in at minimum two intersection
points in the gap profile (or zero values crossings for the gap contrast,
respectively). Thus, the periodicity of the moonlet motion is well refelcted by
the propeller.
\item[ii)] $T_m = T_{gap}:$ At maximum two (at minimum one) intersection points in
the azimuthal gap profile can be identified, depending on the libration phase of
the moonlet at the observation time. 
\item[iii)] $T_m > T_{gap}:$ The slow outward and inward migration of the
moonlet results in at maximum one (minimum zero) observable intersection points.
\end{itemize}

In order to identify a librational motion of the moonlet, the libration
period needs to be smaller than the memory time $T_m < T_{gap}$. 
For larger libration periods the gap contrast will only change its sign 
at the turning points of the radial moonlet libration. In between no change in 
sign will be visible and therefore the moonlet libration might be 
misinterpreted as a radial drift.

\subsection{Observability of the Asymmetry}
In order to measure the asymmetry of the propeller structure, our analysis has
shown, that the azimuthal profiles are the most favorable to consider for 
processing Cassini ISS images.
Especially for larger azimuthal distances to the moonlet ($\Delta y > \pm
\SI{100}{\hill}$), already small changes in the gap minima can result in large 
differences in the gap lengths.

However, in case of high-resolution images, such as UVIS scans, small 
variations in the radial gap profile can be resolved, allowing to study the 
asymmetry even in those profiles.

Nevertheless, in all cases both, inner and outer gap, need to be recorded.

\subsection{Asymmetry of the Propeller \bleriot}
Our predictions yield, that an asymmetry in the gap contrast should be 
visible by about \SI{6.3}{\percent}, which still is larger than the noise 
level for the unperturbed propeller. The gap width differences comparing inner 
and outer gap structure should be about \SI{0.5}{\hill} and the gap lengths are 
expected to differ by \SI{125}{\hill}. 
These variations should be still detectable, but depend on the libration phase
of the central moonlet at the time of observation. 

Due to the large libration period of \bleriot{} at maximum one zero value 
crossing for the gap contrast can be expected, which makes it impossible to 
judge whether the moonlet is librating or migrating considering one single 
observation. An analysis of several images of \bleriot{} at different 
times would be necessary to search for moving patterns (changes in the sign and
amplitudes) in the gap contrast along the azimuth.

\subsection{Asymmetry of other Propellers}
Although the asymmetry for \bleriot{} might be rather small, the asymmetry 
for the other giant trans-Encke propellers might be easier to detect. 
As an example, the propeller structures Earhart and Santos Dumont are smaller
than \bleriot{} but show similar residual amplitudes
\citep{Tiscareno2010ApJL,Spahn2018Book}, resulting in larger radial 
amplitudes for an assumed moonlet libration. Further, due to the smaller size
of these moonlets, their induced propeller wings' azimuthal extent is smaller 
by $M_m/\nu_0 \propto h^3/\nu_0 \propto L/\nu_0$ \citep{Spahn2000AAP}, allowing 
to record both propeller wings at the same time with one observation.

\subsection{Propeller-Moonlet Interactions}
The formation of an asymmetric propeller structure due to the motion of the
moonlet causes a non-balancing force of the ring material on the
moonlet. Calculating the force of the ring material on the moonlet we find, 
that it is dominated by the azimuthal amplitude of the moonlet 
\citep{Seiler2017ApJL} and follows the moonlet motion with a phase shift 
according to the Kepler shear. Thus, our simulations agree with the simple 
propeller-moonlet interaction model suggested by \citet{Seiler2017ApJL}.

\subsection{Outlook}
Although the implementation of a librating moonlet is a simplification, our 
simulation however gives an insight of how the propeller structure reacts to 
perturbations and how the resulting asymmetry will look like. Further, our 
simulations have shown, that the retardation along the azimuth is the main 
mechanism to make the asymmetry measurable in the images. Therefore, even 
for an underlaying stochastic migration of the central moonlet, an asymmetry 
might form. The observability of the resulting asymmetry then will depend on 
the collision frequency and the memory time of the gap.

However, in order to build a fully consitent model, as a next step, the 
moonlet needs to be simulated in a box where it is allowed to move freely. In
this way, a stochastic moonlet motion can be simulated as well, which can be
directly compared to the unperturbed and librating moonlet.

Further, allowing the moonlet to move freely within the simulation box will 
show, whether the back-reaction of the ring material is able to conserve an 
initial libration of the moonlet, which will be the final consistency test for 
the suggested propeller-moonlet interaction model.

Our introduction of the gap contrast allows to directly observe the propagation
of the perturbation by the moonlet motion along increasing azimuth. It turned 
out that the perturbation by the moonlet's motion appears and behaves like 
a moving wave package which gets smeared along the azimuth due to the 
damping by the ring material. This permits to study the dispersion 
relation of the perturbation and further to estimate the viscosity of the 
surrounding ring material from the images.

Our simulation code uses an isothermal model for the ring environment. This 
might result in larger viscosity values in the close vicinity of the moonlet. 
For a consistent model, the temperature needs to be considered in the simualtions 
as well by including the energy equation. This might have an additional effect 
on the observability of the asymmetry.


\acknowledgments
This work has been supported by the Deutsche Forschungsgemeinschaft 
( Sp 384/28-1,2, Ho 5720/1-1) and the Deutsches Zentrum f{\"u}r Luft-und Raumfahrt
(OH 1401).

\listofchanges

\end{document}